\documentclass[fleqn,usenatbib]{mnras}

\usepackage{newtxtext,newtxmath}
\usepackage[T1]{fontenc}

\DeclareRobustCommand{\VAN}[3]{#2}
\let\VANthebibliography\thebibliography
\def\thebibliography{\DeclareRobustCommand{\VAN}[3]{##3}\VANthebibliography}

\usepackage{graphicx}
\graphicspath{./figs/}
\usepackage{amsmath}
\usepackage{natbib}
\usepackage{bm}
\usepackage{color}
\usepackage{natbib}
\usepackage{amsfonts}
\usepackage{multirow}
\usepackage{aas_macros}
\usepackage{threeparttable}
\usepackage{url}
\usepackage{xcolor}
\usepackage{twoopt}
\usepackage{orcidlink}
\newcommandtwoopt{\myfrac}[4][0pt][0pt]{\genfrac{}{}{}{}{\raisebox{#1}{$#3$}}{\raisebox{-#2}{$#4$}}}
\title[A tale of two mass functions]{Discovering gravitationally lensed gravitational waves: predicted rates, candidate selection, and localization with the Vera Rubin Observatory}

\author[G.\ P.\ Smith et al.]
{Graham P.\ Smith\orcidlink{0000-0003-4494-8277},$\!^1$\thanks{E-mail: gps@star.sr.bham.ac.uk}
Andrew Robertson\orcidlink{0000-0002-0086-0524},$\!^2$
Guillaume Mahler\orcidlink{0000-0003-3266-2001},$\!^{3,4}$
Matt Nicholl\orcidlink{0000-0002-2555-3192},$\!^{1,5}$
Dan Ryczanowski\orcidlink{0000-0002-4429-3429},$\!^1$
\newauthor
Matteo Bianconi\orcidlink{0000-0002-0427-537},$\!^1$
Keren Sharon\orcidlink{0000-0002-7559-0864},$\!^6$
Richard Massey\orcidlink{0000-0002-6085-3780},$\!^4$
Johan Richard\orcidlink{0000-0001-5492-1049},$\!^7$
Mathilde Jauzac\orcidlink{0000-0003-1974-8732}.$\!^{3,4,8,9}$
\\
\\
$^1$ School of Physics and Astronomy, University of Birmingham, Edgbaston, B15 2TT, UK\\
$^2$ Jet Propulsion Laboratory, California Institute of Technology, 4800 Oak Grove Dr., Pasadena, CA 91109, USA\\
$^3$ Centre for Extragalactic Astronomy, Department of Physics, Durham University, South Road, Durham DH1 3LE, UK\\
$^4$ Institute for Computational Cosmology, Department of Physics, Durham University, South Road, Durham DH1 3LE, UK\\
$^5$ Institute for Gravitational Wave Astronomy, University of Birmingham, Birmingham B15 2TT, UK\\
$^6$ Department of Astronomy, University of Michigan, 1085 S. University Ave, Ann Arbor, MI 48109, USA\\
$^7$ Univ Lyon, Univ Lyon1, Ens de Lyon, CNRS, Centre de Recherche Astrophysique de Lyon UMR5574, 69230, Saint-Genis-Laval, France\\
$^8$ Astrophysics Research Centre, University of KwaZulu-Natal, Westville Campus, Durban 4041, South Africa\\
$^9$ School of Mathematics, Statistics \& Computer Science, University of KwaZulu-Natal, Westville Campus, Durban 4041, South Africa
}

\date{Accepted 2023 January 9. Received 2022 December 8; in original form 2022 April 28.}
\pubyear{2023}

\defcitealias{GWTC3}{LVK2021}

\begin{document}
\label{firstpage}
\pagerange{\pageref{firstpage}--\pageref{lastpage}}
\maketitle

\newcommand\Mpc{\ensuremath\mathrm{Mpc}}
\newcommand\Om{\ensuremath{\Omega_\mathrm{M}}}
\newcommand\kms{\ensuremath\mathrm{km\,s^{-1}}}
\newcommand\ls{\ensuremath{\hbox{\rlap{\hbox{\lower4pt\hbox{$\sim$}}}\hbox{$<$}}}}
\newcommand\gs{\ensuremath{\hbox{\rlap{\hbox{\lower4pt\hbox{$\sim$}}}\hbox{$>$}}}}
\newcommand\dl{\ensuremath{D_\mathrm{L}}}
\newcommand\ds{\ensuremath{D_\mathrm{S}}}
\newcommand\dls{\ensuremath{D_\mathrm{LS}}}
\newcommand\dlC{\ensuremath{D_\mathrm{L}^{\mathrm C}}}
\newcommand\dsC{\ensuremath{D_\mathrm{S}^{\mathrm C}}}
\newcommand\dlsC{\ensuremath{D_\mathrm{LS}^{\mathrm C}}}
\newcommand\zs{\ensuremath{z_{\rm S}}}
\newcommand\zl{\ensuremath{z_{\rm L}}}
\newcommand\valpha{\ensuremath{\bm{\alpha}}}
\newcommand\vbeta{\ensuremath{\bm{\beta}}}
\newcommand\vtheta{\ensuremath{\bm{\theta}}}
\newcommand\mup{\ensuremath{\mu_{\rm p}}}
\newcommand\thE{\ensuremath{\theta_{\rm E}}}
\newcommand\kE{\ensuremath{\kappa_{\rm E}}}
\newcommand\etaE{\ensuremath{\eta_{\rm E}}}
\newcommand\tz{\ensuremath{\widetilde{z}}}
\newcommand\tD{\ensuremath{\widetilde{D}}}
\newcommand\tM{\ensuremath{\widetilde{\mathcal{M}}}}
\newcommand\tm{\ensuremath{\widetilde{m}}}
\newcommand\mMmin{\ensuremath{\mathcal{M}_{\rm min}}}
\newcommand\mMmax{\ensuremath{\mathcal{M}_{\rm max}}}
\newcommand\mmin{\ensuremath{m_{\rm min}}}
\newcommand\mmax{\ensuremath{m_{\rm max}}}
\newcommand\mbr{\ensuremath{m_{\rm br}}}
\newcommand\pdet{\ensuremath{p_{\rm det}}}
\newcommand\mR{\ensuremath{\mathcal{R}}}
\newcommand\mM{\ensuremath{\mathcal{M}}}
\newcommand\mD{\ensuremath{\mathcal{D}}}
\newcommand\Ndet{\ensuremath{N_{\rm det}}}
\newcommand\Dmax{\ensuremath{D_{\rm max}}}
\newcommand\zmax{\ensuremath{z_{\rm max}}}
\newcommand\Narr{\ensuremath{N_{\rm arr}}}
\newcommand\zp{\ensuremath{z_{\rm p}}}
\newcommand\Dt{\ensuremath{\Delta t}}
\newcommand\Dtheta{\ensuremath{\Delta\theta}}
\newcommand\sigcrit{\ensuremath{\Sigma_{\rm crit}}}
\newcommand\thA{\ensuremath{\theta_{\rm A}}}
\newcommand\thB{\ensuremath{\theta_{\rm B}}}
\newcommand\mfobs{\ensuremath{m_{\rm obs}}}
\newcommand\mfint{\ensuremath{m_{\rm int}}}
\newcommand\pgap{\ensuremath{p_{\rm gap}}}
\newcommand{\thetap}{\ensuremath{\theta_+}}
\newcommand\thetam{\ensuremath{\theta_-}}
\newcommand\tauI{\ensuremath{\tau^{\rm I}}}
\newcommand\tauS{\ensuremath{\tau^{\rm S}}}
\newcommand\phip{\ensuremath{\phi_{\rm p}}}
\newcommand\sigmam{\ensuremath{\sigma_{\rm m}}}
\newcommand\muth{\ensuremath{\mu_{\rm th}}}

\begin{abstract}
{Secure confirmation that a gravitational wave (GW) has been gravitationally lensed would bring together these two pillars of General Relativity for the first time. This breakthrough is challenging for many reasons, including: GW sky localization uncertainties dwarf the angular scale of gravitational lensing, the mass and structure of gravitational lenses is diverse, the mass function of stellar remnant compact objects is not yet well constrained, and GW detectors do not operate continuously. We introduce a new approach that is agnostic to the mass and structure of the lenses, compare the efficiency of different methods for lensed GW discovery, and explore detection of lensed kilonova counterparts as a direct method for localising candidates. Our main conclusions are: (1) lensed neutron star mergers (NS-NS) are magnified into the ``mass gap'' between NS and black holes, therefore selecting candidates from public GW alerts with high mass gap probability is efficient, (2) the rate of detectable lensed NS-NS will approach one per year in the mid-2020s, (3) the arrival time difference between lensed NS-NS images is $1\,\rm sec\lesssim\Delta t\lesssim1\,year$, and thus well-matched to the operations of GW detectors and optical telescopes, (4) lensed kilonova counterparts are faint at peak (e.g.\ $r_{\rm AB}\simeq24-26$ in the mid-2020s), fade quickly ($d<2\,\rm days$), and are detectable with target of opportunity observations with large wide-field telescopes. For example, just $\lesssim0.25$ per cent of Vera C.\ Rubin Observatory's observing time will be sufficient to follow up one well-localized candidate per year. Our predictions also provide a physically well-defined basis for exploring electromagnetically the exciting new ``mass gap'' discovery space.}
\end{abstract}


\section{Introduction}\label{sec:intro}

Gravitational waves (GWs) are magnified, deflected, and delayed as they traverse gravitational fields, just like the more familiar gravitational lensing of light from distant galaxies and quasars \citep{Ohanian1974,Wang1996,Nakamura1998}, pointing to exciting opportunities to detect multiple gravitationally magnified images of a distant GW source. Detection of gravitationally lensed GWs will be the first opportunity to combine these two pillars of Einstein's General Relativity in a single experiment, leading to novel tests of General Relativity and general metric gravity theories \citep[e.g.][]{Mukherjee2020weak,Mukherjee2020galaxy,Goyal2021,Finke2021}, measurements of the Hubble constant and tests of the Friedmann-LeMa\^itre-Robertson-Walker metric \citep[e.g.][]{Sereno2011,Liao2017,Cao2019,Li2019,Hannuksela2020,Hou2020}, testing the speed of GW propagation \citep[e.g.][]{Baker2017,Collett2017,Fan2017}, and constraints on intermediate mass black holes (BH) and primordial BHs via their microlensing signal \citep{Lai2018,Diego2020,Oguri2020,Wang2021,Urrutia2022}. These studies set a broad expectation that gravitationally lensed GWs is the science of the late 2030s and 2040s when the 3rd generation of GW detectors will begin operations, generally focus on lensed binary black hole (BH-BH) mergers over lensed binary neutron star (NS-NS) mergers, and tend to assume that the only relevant population of lenses is massive early-type galaxies. We discuss these three themes in turn below.

In the more familiar realm of gravitational lensing of electromagnetic (EM) radiation, early detections of gravitationally lensed high redshift galaxies, some of which were the redshift record breakers of their day \citep[e.g.][]{Soucail87a,Lynds1989,Mellier91,Franx97,Ellis01,Kneib04b}, pre-dated systematic studies of large samples of lensed galaxies  by several decades \citep[e.g.][]{Atek2015,McLeod2016,Ishigaki2018,Bhatawdekar2019}. {Current progress towards} the first detection and early science of gravitationally lensed GWs therefore resonates with the status of lensed high redshift galaxies in the mid-1980s and detection of lensed quasars a little earlier than that \citep{Walsh1979} -- i.e.\ the years leading up to the first detection. {Likewise,} systematic study of large samples of lensed GWs with next generation GW instruments in the 2030s and 2040s will resemble the status of lensed galaxy and lensed quasar studies of the last decade. 

A growing suite of predictions are consistent with a rate of detection of lensed GWs of around one per year in the coming years {with the current generation of GW detectors} \citep{Oguri2018,Oguri2019,Li2018,Ng2018,Smith2019lsst,Wierda2021,Xu2022}. {Echoing the progress driven by the emergence of} charge coupled devices (CCDs) in optical astronomy in the 1980s, several groups have therefore been considering whether any of the gravitationally lensed GWs detected by current instruments are lensed \citep[e.g.][]{Smith2018,Smith2019,Broadhurst2018,Hannuksela2019,Singer2019,Dai2020,GW190521prop,McIsaac2020,Liu2021,LVlens2021,Diego2021,Bianconi2022}. {In turn, this is motivating development of techniques to identify candidate lensed GWs  \citep[e.g.][]{Oguri2018,Li2018,Ng2018,Smith2019lsst,Hannuksela2020,Robertson2020,Ryczanowski2020,Ryczanowski2022,Wempe2022,Wierda2021,Xu2022,Yang2022}, and joint investigations of lensed GWs and the stocahastic GW background \citep{Buscicchio2020b,Buscicchio2020a,Mukherjee2021stochastic}}.

It is inarguable that multi-messenger detection of a lensed GW will be transformational, i.e.\ detection of a lensed NS-NS merger and its lensed EM counterpart. This is because the EM detection will localise the GW source to $\lesssim1\,\rm arcsec$ accuracy with ground-based data, and potentially $\simeq10\,\rm milliarcsec$ accuracy with space-based observations. This will be the lensing analogue of the localisation of GW170817 to a specific location in the outskirts of its host galaxy NGC4993 via the detection of its optical/infrared and short gamma-ray burst (SGRB) counterpart \cite[][and references therein]{GW170817-MMA,MarguttiChornock2021}. Such accurate localisation would eliminate reliance on statistical arguments for or against several poorly localised BH-BH detections being images of the same lensed source and several different galaxies being the lens and/or the host \citep[e.g.][]{Hannuksela2020,Wempe2022}. {Moreover, despite intriguing candidates \citep[e.g.][]{Graham2020}, there are no confirmed EM counterparts to BH-BH mergers available to help with localisation.} Whilst the detection rates of GW signals from lensed NS-NS mergers are expected to be lower than for lensed BH-BH mergers, preliminary estimates indicate that multi-messenger detection of lensed NS-NS mergers will be feasible in the 2020s \citep{Smith2019lsst,Oguri2019}. 

Detection of multiple images of the lensed EM counterpart will also expand the science of lensed GWs, for example into the chemical enrichment history of the universe and the physics of kilonovae. For example, detection of the trailing image of the lensed kilonova will potentially probe the rising portion of the kilonova in the rest-frame ultraviolet (UV). This would shed light on arguably the biggest unsolved mystery of GW170817, namely the physics responsible for the very bright blue early component of its emission \citep[e.g.][]{Arcavi2018,Piro2018,Nicholl2021}. The feasibility of such measurements will depend on the arrival time difference between the images and the reliability {of any predictions that are required}, as highlighted by resolved and  confirmed detections of strongly lensed supernovae  \citep{Kelly2015,Treu2016,Jauzac2016,Goobar2017,Goobar2022,More2017}.

The optical depth to strong lensing (defined here as large lens magnification, $\mu>10$) spans galaxies, groups and clusters of galaxies \citep{Hilbert2008}, with the latest calibration being  consistent with lenses in each decade of mass spanning $10^{12}\le M_{200}\le10^{15}\,\rm M_\odot$ contributing approximately equally to the optical depth \citep{Robertson2020}. The scaling of lensing probability with magnification in the strong lensing regime is also well defined analytically as $dp/d\mu\propto\mu^{-3}$ and -- importantly -- independent of the lens mass \citep{Blandford1986}. Consequently \citeauthor{Hilbert2008} and \citeauthor{Robertson2020}'s results persist at all magnifications of $\mu>2$. 

The relationship between lens magnification and image multiplicity is less well defined, as it depends on the relationship between the mass and the density profile slope of lenses. The basic idea was sketched by \cite{Turner1984} and is still valid today  \citep[e.g.][]{Smith2001a,Gavazzi03,Sand04,Smith2009,Richard2010,Umetsu2016,Fox2021}. Briefly, galaxy clusters tend to have shallower density profiles than individual galaxies and therefore the former are less efficient than the latter at forming multiple images of background objects at low magnification, typically $2<\mu\lesssim10$. However, clusters still contribute their ``fair share'' of the optical depth at low magnification, they just do so by forming a single image. This is important in the context of predicting rates of and searching for gravitationally lensed GWs. In particular, a currently un-constrained and potentially significant fraction of the optical depth to lens magnifications of $2<\mu\lesssim10$ does not contribute to multiple image formation. Conversely, as highlighted by resolved confirmed strongly lensed supernova discoveries \citep{Kelly2015,Goobar2017,Goobar2022}, searching for high magnification strong lensing requires consideration of lenses spanning galaxies, groups and clusters and not just a focus on clusters. 

Decisive progress towards robust and unchallenged detection of a gravitationally lensed GW rests on identifying candidates and localizing them to a lensed host galaxy. Searches for candidates have so far identified either pairs of GW detections with sky localizations and waveforms that are consistent with being lensed images of the same source, and/or individual GW sources that appear to have an anomalous mass relative to an assumed underlying mass function \citep{Smith2018,Smith2019,Hannuksela2019,Singer2019,Dai2020,McIsaac2020,GW190521prop,Diego2021,Liu2021,LVlens2021,Diego2021,Bianconi2022}. {The former approach relies on the GW detectors collecting data of the required quality when $\ge2$ of the lensed images arrived at Earth, and the latter approach is sensitive to the assumed mass function. For example, \cite{Broadhurst2018,Broadhurst2019,Broadhurst2022} argue that if the true stellar remnant BH mass function in the universe peaks at $M_{\rm BH}\simeq10\,\rm M_\odot$ (i.e. well-matched to the BHs thus far detected in the Milky Way), then the GW detections apparently from systems comprising BH in the range $M_{\rm BH}\gtrsim20\,\rm M_\odot$ are dominated by lensed events.} 

{Efficient selection of candidate lensed GWs therefore faces several important challenges.} Firstly, how to define anomalous when the underlying mass function of sources is not known a priori? Secondly, how efficient is searching for multiple detections in a rapidly growing catalogue of GW detections from the LIGO, Virgo and KAGRA (LVK) detectors that are dominated by sources with 90 per cent credible sky localisation uncertainties of $\Omega_{90}>100\,\rm degree^2$ \citep{Petrov2021}? Moreover, the GW detectors rarely operate uninterrupted for more than a few hours and are offline for technology upgrades for timescales of $\gs1\,\rm year$.

In this article we develop a new magnification-based approach to predicting rates of gravitationally lensed GWs that naturally integrates strong lenses of all masses and thus allows us to calibrate both the rates of lensed GW detections, the typical magnifications that they suffer, and their location in the mass-distance plane. We apply this approach to lensing of both BH-BH and NS-NS mergers and combine our magnification-based predictions with time delay theory to identify the typical arrival time differences for these sources and lenses of different mass and structure. This enables us to combine our predictions in the mass-distance plane with expected arrival time differences to develop efficient strategies for identifying and localizing candidate lensed GWs. This leads to predicting the lightcurves of lensed kilonova counterparts of lensed NS-NS mergers and considering their detectability with the Vera C. Rubin Observatory (Rubin). 

We review the theory and phenomenology of strong lenses and develop {our new magnification-based approach} in Section~\ref{sec:lensing}, then describe the GW source populations that we consider in Section~\ref{sec:gw}, before applying the former to the latter and making our predictions in Section~\ref{sec:pred:gw}. We combine our lensed GW predictions with \citeauthor{Nicholl2021}'s (\citeyear{Nicholl2021}) kilonova lightcurve models to predict lensed kilonova lightcurves and assess the feasibility of detection with Rubin Target of Opportunity observations in Section~\ref{sec:observing}, and close with a summary in Section~\ref{sec:summary}. We assume a flat cosmology with $H_0=67.9\,\kms\,\Mpc^{-1}$, $\Om=0.3065$ \citep{Planck15cos}. All magnitudes are stated in the AB system.

\section{Gravitational lensing theory and phenomenology}\label{sec:lensing}

We begin with an overview of the main take away points in Section~\ref{sec:lensing:overview}. For readers interested in more detail, we outline some key elements of gravitational lensing theory in Section~\ref{sec:lensing:theory}, before motivating our focus on gravitational magnification in Section~\ref{sec:lensing:versus}, justifying our model for optical depth in Section~\ref{sec:lensing:odsl}, and explaining how we predict arrival time difference for magnified image pairs in Section~\ref{sec:lensing:time}.

\subsection{Overview}\label{sec:lensing:overview}

We consider strong gravitational lensing of GW and EM signals from mergers of compact object binaries, where the lenses are galaxies, groups and clusters of galaxies that are serendipitously located along the line of sight to the merger. {We ignore any microlensing by stars or compact objects that may be located close to the critical curves of the ``macrolens'' -- i.e. the galaxy/group/cluster lens. Whilst microlensing can be important, especially in high magnification regions of very dense macrolenses \citep[e.g.][]{Diego2019micro,Diego2019extreme,Meena2020,Mishra2021}, it is beyond our current aim of establishing baseline predictions of, and follow-up strategies for, candidate gravitationally lensed GWs. As such our approach is likely to yield lower limits on the detectable rates of lensed GWs.}

The geometrical nature of gravitational lensing and the growth of large scale structure in the universe result in strong lensing being dominated by lenses at redshifts of $\zl\simeq0.2-0.6$ and sources at redshifts of $\zs\,\gs\,1$. The typical comoving distance travelled by a signal from a strongly lensed GW source is therefore expected to be {$\gs7\,\rm Gpc$ ($\gs10^{26}\,\rm m$)}. This is three orders of magnitude larger than the comoving virial radius of the most massive galaxy clusters in the observable universe \citep{Bianconi2021}. It is therefore valid to adopt the so-called thin lens approximation. 

The distance to strongly lensed GW sources also dwarfs their physical size ($\lesssim10^3\rm m$), allowing us to treat them as point sources of EM and gravitational radiation. This allows us to ignore finite source size effects when considering large lens magnifications. Despite the wavelength of GWs ($\lambda_{\rm GW}\lesssim10^7\rm m$ for LVK) being much longer than EM radiation, the former is still many orders of magnitude  smaller than the physical scale of the lenses (Einstein radii corresponding to physical scales of $\gtrsim10^{19}\rm m$, {i.e.\ $\gtrsim300\,\rm pc$, and thus $\gtrsim0.1\,\rm arcsec$ at $z\simeq0.2$}). For the purposes of our calculations we are therefore safe to treat both GWs and EM radiation in the geometric optics limit and thus ignore wave effects.

We predict the rates and properties of gravitationally lensed GWs and their EM counterparts as a function of gravitational lens magnification. As discussed in Section~\ref{sec:lensing:versus}, this is the least model-dependent approach to predicting the ensemble of lensed detections that are feasible with GW detectors and wide-field optical telescopes at the current time. In particular, we side-step the challenge of solving the lens equation (Equation~\ref{eqn:lensequation}) and the requirement to assume a form for the deflection potential that this would entail. The main advantage of our approach, beyond computational speed, is that not assuming anything about the structure of the lens population means that our predictions are not biased to any type of lens and are thus relevant across the full range of lens mass ($10^{12}\lesssim M_{200}\lesssim10^{15}\,\rm M_\odot$; \citealt{Robertson2020}). The main disadvantage is that the efficiency of multiple image production at low lens magnification ($2<\mu\lesssim10$) depends on lens mass, with individual galaxy-scale lenses (relatively steep density profiles) being more efficient than group- and cluster-scale lenses (relatively shallow density profiles) at low magnification. Care is therefore needed at $\mu\lesssim10$, because some of the predicted magnified detections will be single images. 

We adopt a threshold magnification of an individual GW detection of $\mu>\muth=2$. This threshold corresponds to a bias by a factor of $>\sqrt{\muth}=1.4$ in the luminosity distance to a lensed source if the magnification is ignored (Equation~\ref{eqn:mubias}), i.e.\ comparable with or larger than the typical 90 per cent posterior credible intervals on luminosity distance in LVK's GW data analysis. This threshold also marks the onset of multiple image formation by a lens -- such as an isolated early type galaxy -- that is well-described by {an isothermal density profile.} It is however important to note that the fainter of two images formed by an isothermal lens is demagnified, {i.e.\ $0\le|\mu_-|\le1$, if the magnification of the brighter image is in the range $2\le\mu_+\le3$} (Equation~\ref{eqn:musis}). {The formation of multiple images therefore does not automatically imply the detectability of multiple images, although detections at $\mu\gtrsim10$ will be dominated by multiply-imaged systems (Figure~\ref{fig:foxetal+odsl} and Section~\ref{sec:lensing:versus}), and be accompanied by other similarly magnified images independent of the structure of the lens. }

We use analytic expressions to convert the predicted magnifications of gravitationally lensed GWs in to predictions of the time since / until another signal from the same source arrived / will arrive at Earth (Equations~\ref{eqn:tpseudo}~\&~\ref{eqn:tfold}). These expressions are based on two common scenarios for the formation of bright highly magnified image pairs, and are shown to capture the time delay behaviour of the systems that have been discovered to date (Figure~\ref{fig:dt:verify}). {The expressions capture the dependence of time delay on both the mass and the structure of the lens, bearing in mind that group/cluster-scale lenses} tend to be denser and flatter than galaxy-scale lenses at their respective Einstein radii (Figure~\ref{fig:foxetal+odsl}). The details of the time delay and its dependence on lens properties are described in Section~\ref{sec:lensing:time} and the supporting derivations are presented in Appendix~\ref{app:derivation}. 

\subsection{Lensing fundamentals}\label{sec:lensing:theory}

We consider an infinitely thin mass distribution at redshift, $\zl$, {and distant sources at redshift $\zs>\zl$}. The true and  unmeasurable positions of the sources on the celestial sphere are $\vbeta$ and the observed positions are $\vtheta$. Following precepts laid down by \cite{Schneider1985}, \cite{Blandford1986}, and others, the time for a lensed signal to travel from the source to the observer differs from the travel time in the absence of lensing by an amount that depends on a dimensionless scalar field called the Fermat potential, $\tau$:
\begin{equation}
    c\,t=\mathcal{D}\,\tau(\vtheta,\vbeta),
    \label{eqn:ct}
\end{equation}
where $\mD=\dlC\dsC/\dlsC$, $\dlC$ and $\dsC$ are the comoving distances from the observer to the lens and the source respectively, and $\dlsC=\dsC-\dlC$. The Fermat potential comprises a geometrical term that depends on image and source positions, and the deflection potential, $\psi$, that describes the so-called \citet{Shapiro1964} delay:
\begin{equation}
    \tau(\vtheta,\vbeta)=\frac{(\vtheta-\vbeta)^2}{2}-\psi(\vtheta).
  \label{eqn:fermat}
\end{equation}
The deflection potential satisfies the Poisson equation,
\begin{equation}
    \nabla^2\psi=2\kappa,
    \label{eqn:poisson}
\end{equation}
where $\kappa$ is the projected density of the lens in units of the critical surface mass density, $\sigcrit$,
\begin{equation}
  \kappa(\vtheta)\equiv\frac{\Sigma(\vtheta)}{\sigcrit},
  \label{eqn:kappa_defn}
\end{equation}
where
\begin{equation}
  \sigcrit=\frac{c^2}{4\pi G}\,\frac{\ds}{\dl\dls},
\end{equation}
and $\dl$, $\dls$, and $\ds$ are angular diameter distances.

By Fermat's Principle, the measurable positions of gravitationally lensed sources (commonly referred to as images) are located at the stationary points of the surface described by Equation~\ref{eqn:fermat}. Taking the gradient and setting $\nabla\tau=0$ yields the lens equation:
\begin{equation}
  \vtheta=\vbeta+\nabla\psi(\vtheta)=\vbeta+\valpha(\vtheta),
  \label{eqn:lensequation}
\end{equation}
where $\valpha=\nabla\psi$ is referred to as the deflection angle. In the strong lensing regime multiple images of a distant source that is observed through a gravitational lens occur if Equation~\ref{eqn:lensequation} has multiple solutions, $\vtheta_k$, for a given source position, $\vbeta$. 

The angular offset of multiple images of a distant source from the centre of a lens is often characterized by the Einstein radius of the lens. Formally, this is the radius of a continuous ring-like image of a source that is located precisely on the axis of an axisymmetric lens:
\begin{equation}
  \thE=\left(\frac{4GM}{c^2}\,\frac{\ds}{\dl\dls}\right)^{1/2},
  \label{eqn:thE}
\end{equation}
where $M$ is the projected mass interior to $\thE$. Whilst observed gravitational lenses are only approximately axisymmetric, $\thE$ is a useful approximation for the location of strongly-lensed images of distant sources. It also emphasises the convenience of $\kappa$, because $\kappa\ge1$ is sufficient for the formation of multiple images \citep{Subramanian1986}, and $\langle\kappa(<\thE)\rangle\equiv1$ for axi-symmetric lenses. 

The time delay relative to an unperturbed signal, as defined by Equations~\ref{eqn:ct}~\&~\ref{eqn:fermat}, is not measurable because true source positions are not measurable in the presence of a lens along the line of sight. The measurable quantity is the arrival time difference between two gravitationally lensed images of the same source, which is proportional to the difference between the Fermat potentials traversed by the two signals, here denoted $k=1,2$:
\begin{equation}
  c\,\Delta t_{1,2}=\mD\big[\tau(\vtheta_1,\vbeta)-\tau(\vtheta_2,\vbeta)\big].\\
  \label{eqn:Deltat}
\end{equation}
Equations~\ref{eqn:fermat} and \ref{eqn:Deltat} reveal that the arrival time difference depends on the details of the gravitational potential of the lens, as discussed further in Section~\ref{sec:lensing:time} and Appendix~\ref{app:derivation}.

The apparent size and brightness of gravitationally lensed images differ from the source in the absence of lensing by a factor $\mu$, the lens magnification. This is a geometrical effect described by the determinant of the Jacobian matrix of the coordinate transformation from the ``source plane'', $\vbeta$, to the ``image plane'', $\vtheta$:
\begin{equation}
  \mu=\frac{1}{\det\mathcal{A}},
  \label{eqn:mu}
\end{equation}
where
\begin{equation}
  \mathcal{A}(\vtheta)=\frac{\partial\vbeta}{\partial\vtheta}=\delta_{ij}-\frac{\partial^2\psi(\vtheta)}{\partial\theta_i\,\partial\theta_j}=\delta_{ij}-\psi_{,ij}\label{eqn:jacobian}
\end{equation}
and $\delta_{ij}$ is the identity matrix. Equation~\ref{eqn:jacobian} implies that the magnification factor is the ratio of the solid angle subtended by a magnified image, $\omega_{\rm I}$, to the solid angle that the source would subtend if it were not lensed, $\omega_{\rm S}$:
\begin{equation}
  \mu=\frac{\omega_{\rm I}}{\omega_{\rm S}}.
  \label{eqn:muomega}
\end{equation}
The estimated distance to a magnified source of known luminosity is therefore systematically biased by an amount implied by the inverse square law:
\begin{equation}
  \mu=\left(\frac{D}{\tD}\right)^2\,\Rightarrow\,\tD=\frac{D}{\sqrt{\mu}}\,,
  \label{eqn:mubias}
\end{equation}
where $D$ is the true luminosity distance to the source, and we introduce $\tD$ as the luminosity distance that would be inferred if $\mu=1$ is assumed for a source for which $\mu\ne1$. Phrasing magnification in terms of distance (Equation~\ref{eqn:mubias}) is useful in the context of gravitationally lensed GWs, because the LVK collaboration assumes $\mu=1$ when analyzing their data immediately after detection of a GW, and therefore they infer $\tD$. 

\subsection{Image multiplicity versus magnification}\label{sec:lensing:versus}

\begin{figure*}
  \centerline{
    \includegraphics[width=0.32\hsize,angle=0]{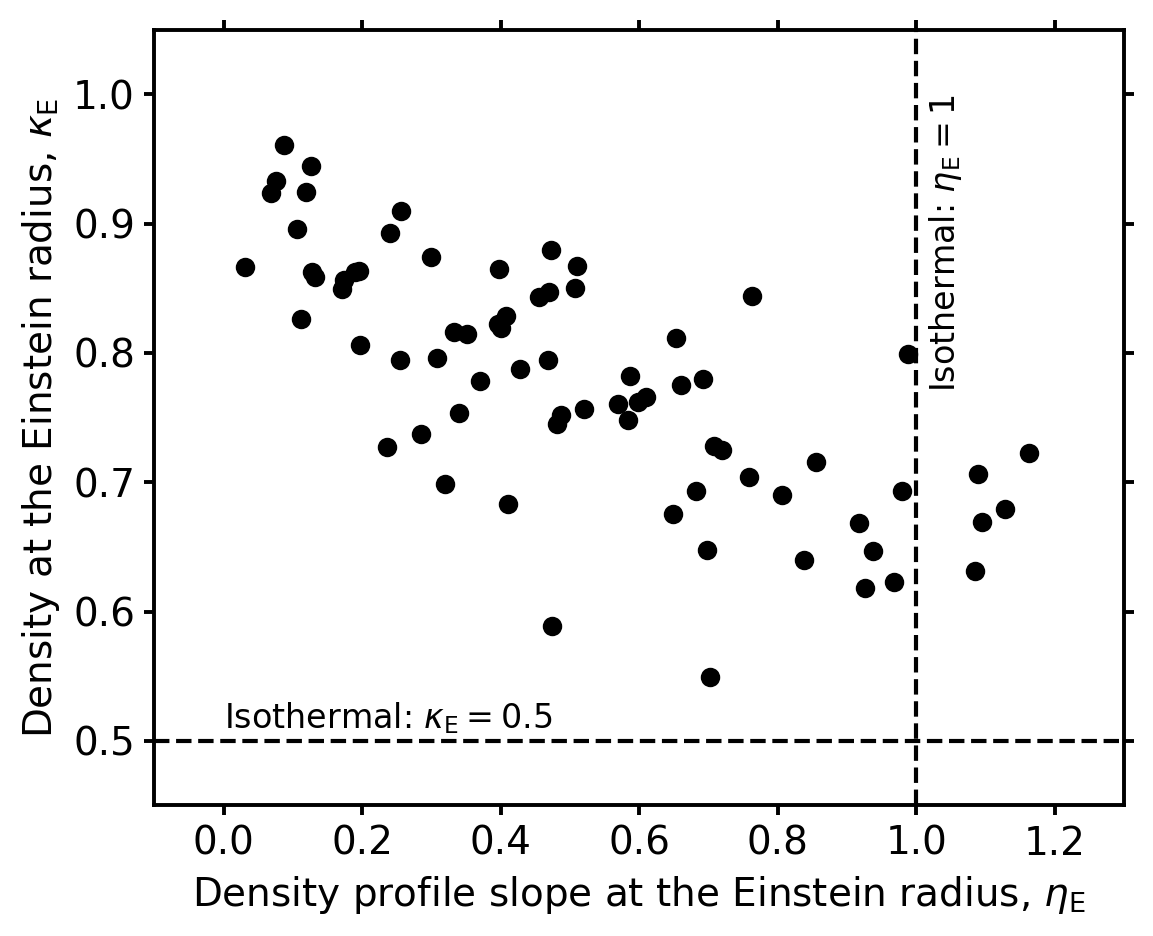}
    \hspace{1mm}
    \includegraphics[width=0.32\hsize,angle=0]{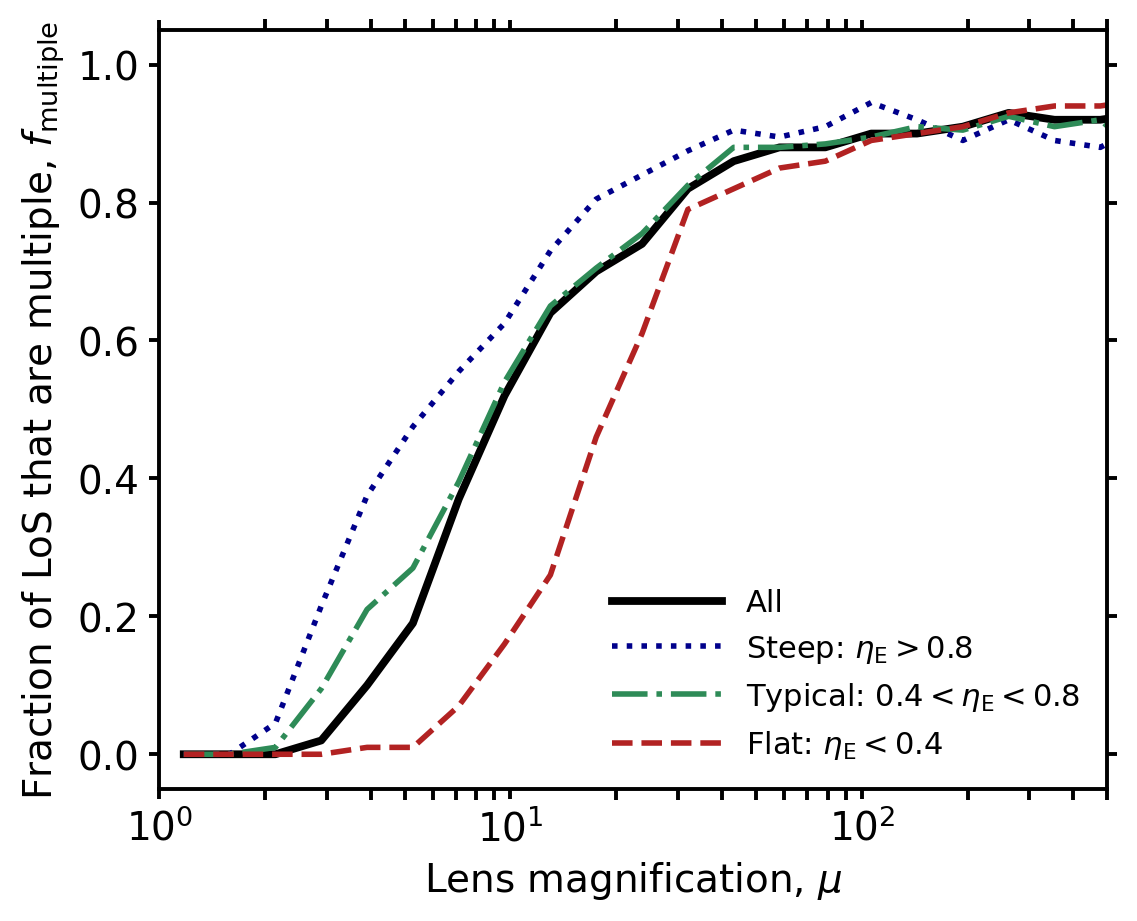}
        \hspace{1mm}
        \includegraphics[width=0.35\hsize,angle=0]    {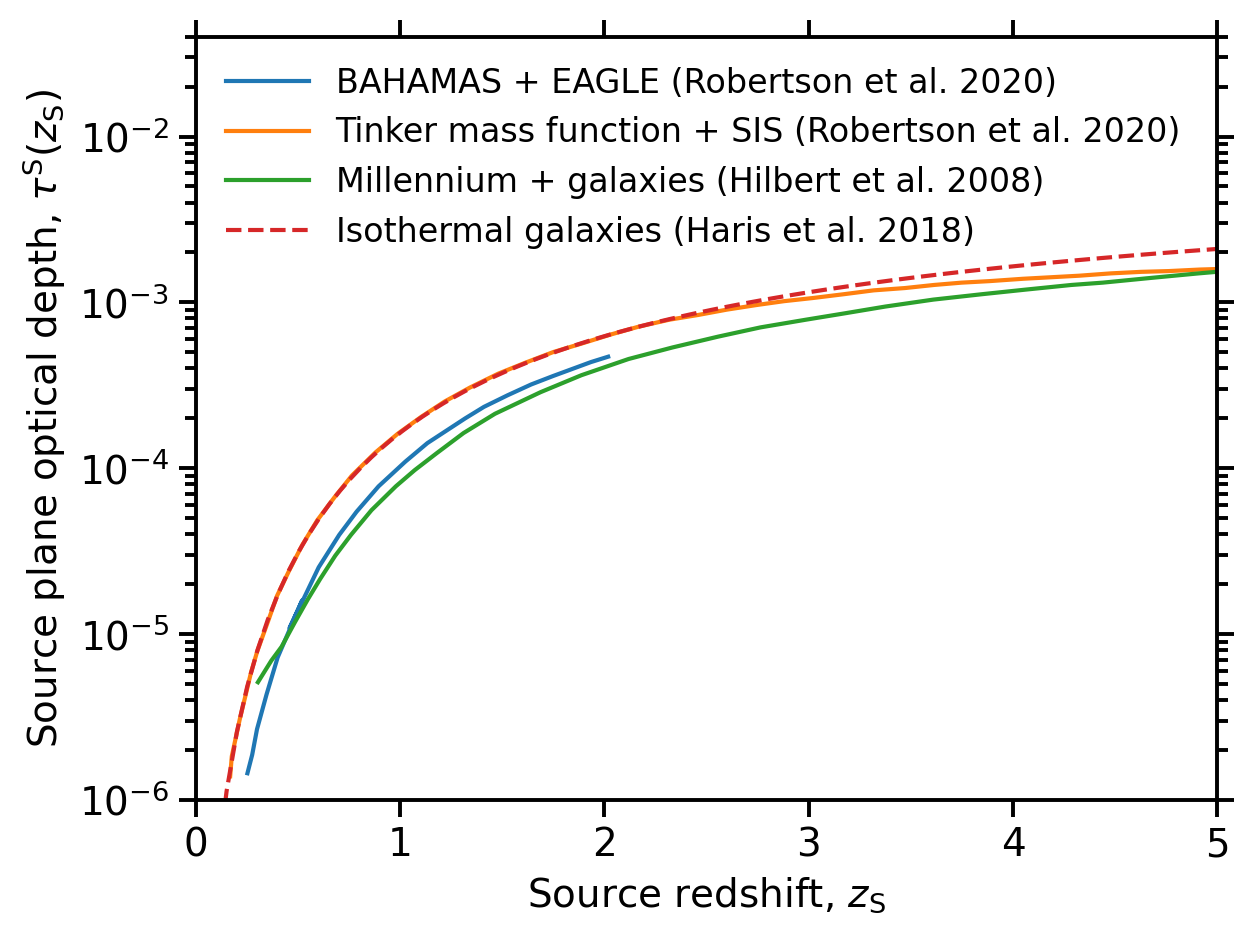}
    }
    \caption{{\sc Left} --- The distribution of density $\kE$ versus density profile slope $\etaE$, measured from the lens models of 79 group- and cluster-scale lenses. Groups and clusters are denser and shallower at their Einstein radius than the SIS model that is a good description of individual galaxy-scale lenses.  {\sc Center} --- The run of $f_{\rm multiple}$ with lens magnification for the  lenses shown in the left panel, where $f_{\rm multiple}$ is the fraction of lines of sight at a given lens magnification that are multiple, i.e. have image multiplicity of greater than one. The threshold magnification at which lenses are efficient at forming multiple images of distant sources is a function of density profile slope. {\sc Right} --- Different predictions of the source plane optical depth as a function of source redshift agree within a factor of $\ls\,2$. The three models shown by solid curves are discussed by \protect\citeauthor{Robertson2020}, and are based on three different approaches to integrating the optical depth to lens magnification over the full dynamic range of the dark matter halo mass function. The BAHAMAS$+$EAGLE curve is based on modern high resolution cosmological hydrodynamical simulations, the Tinker curve combines the analytic form of the \protect\cite{Tinker2008} mass function with the cross-section of the singular isothermal density profile model, and the Millennium curve is based on the semi-analytic approach of \protect\cite{Hilbert2008}, that pastes analytic galaxies in to the dark matter-only Millennium Simulation. The dashed curve, from \protect\cite{Haris2018}, is a prediction of the optical depth to multiple imaging that assumes that all gravitational lenses are individual massive galaxies with isothermal density profiles.}
  \label{fig:foxetal+odsl}
\end{figure*}

In principle, calculating the number of gravitationally lensed images of a given source population involves choosing a deflection field, $\psi$, and using Equation~\ref{eqn:lensequation} to compute all possible image positions, $\vtheta_k$, for all possible source positions, $\vbeta$. However, whilst it is straightforward to compute $\vbeta$ given $\psi$ and $\vtheta_k$, the inverse mapping from $\vbeta$ to an unknown number of $\vtheta_k$ can only be solved analytically for very simple mass distributions, and is time consuming to solve numerically. 

Simple models for the extended mass distributions of gravitational lenses describe the projected density of the lens as a monotonic function of projected physical distance from the centre of the lens, $R=\theta\dl$:
\begin{equation}
  \Sigma(R)\propto R^{-\eta},
  \label{eqn:sigprof}
\end{equation}
with a global power-law slope $\eta>0$ at all radii. The singular isothermal sphere (SIS; $\eta=1$) is the canonical analytically solvable case and is also relevant because models with an isothermal slope (generally singular isothermal ellipses) are appropriate to isolated galaxy-scale lenses \citep[e.g.][]{Koopmans2006}. The projected mass density profile of an isothermal mass distribution is given by:
\begin{equation}
  \Sigma(R)=\frac{\sigma^2}{2GR},
  \label{eqn:sis}
\end{equation}
where $\sigma$ is the line of sight velocity dispersion of the particles in the lens \citep{BinneyTremaine}, and for which it can be shown that the deflection angle is constant and equal to the Einstein radius:
\begin{equation}
  \alpha=\thE=\frac{4\pi\sigma^2}{c^2}\frac{\dls}{\ds}.
\end{equation}

SIS lenses produce pairs of gravitational images with total magnifications of $\mup>2$, where $\mup$ is the sum of the magnifications suffered by the image pair \citep{Turner1984}. The limiting case of $\mup=2$ corresponds to one image (generally denoted as ``$+$'') magnified by $\mu_+=2$ and the other image (generally denoted as ``$-$'') de-magnified such that $\mu_-\rightarrow0$. In general the magnifications suffered by these images are given by:
\begin{equation}
  \mu_+=1+\frac{\thE}{\beta}~{\rm and}~\mu_-=\frac{\thE}{\beta}-1,
  \label{eqn:musis}
\end{equation}
for $\beta\le\thE$. SIS lenses therefore produce two images that are both brighter than the source when $\beta<\thE/2$, which yields $\mup=\mu_++\mu_->4$. Note that the lensing behaviour outlined here relates to the pseudo-caustic catastrophe discussed in Section~\ref{sec:lensing:time}. Realistic isothermal lenses (i.e. perturbed from circular symmetry by the ellipticity of the lens and/or external shear) can also produce additional images, for example quadruply imaged quasars \citep[e.g.][]{Millon2020}.

Group- and cluster-scale lenses are more massive than galaxy-scale lenses (hence larger $\thE$) and typically have density profile slopes at their Einstein radius that are shallower than isothermal. Strong lensing constraints on the density profile slope of group- and cluster-scale lenses have been discussed extensively in the literature dating back to early theoretical work by \cite{Turner1984} and spanning more recent observational studies of increasingly large samples \citep[e.g.][]{Smith2001a, Gavazzi03, Sand04, Smith2009, Richard2010, Umetsu2016, Fox2021}. We illustrate this in Figure~\ref{fig:foxetal+odsl}, based on 79 cluster lens models assembled from \citet{Richard2010}, \citet{Fischer2019} and \citet{Fox2021}, including some models from the SGAS collaboration \citep{Sharon2020}, and the RELICS collaboration \citep{Coe2019}. It is clear that clusters are both denser (relative to $\sigcrit$) and flatter than the isothermal model that is typical of galaxy-scale lenses, where we introduce $\kappa_{\rm E}$ and $\eta_{\rm E}$ as the density and density profile slope respectively at the Einstein radius. For reference, the median cluster in this sample is at a redshift of $z=0.375$, with $\thE=10''$, $\kE=0.78$, and $\etaE=0.48$, where $\thE$ and $\kE$ are stated for a source redshift of $z=1.6$ as the typical source redshift found in Section~\ref{sec:pred:gw}.

We show the implications of the structure of cluster-scale lenses for image multiplicity as a function of magnification in the central panel of Fig.~\ref{fig:foxetal+odsl}. The magnification threshold above which cluster-scale lenses produce multiple images is larger than the threshold relevant to SIS lenses discussed above, and is a function of the profile slope, $\etaE$. The magnification threshold for an individual image (i.e.\ not the sum over an image pair) at which $f_{\rm multiple}\ge0.5$ for the steepest cluster-scale lenses ($\etaE>0.8$) is $\mu\simeq6$, and for the shallowest cluster-scale lenses ($\etaE<0.4$) is $\mu\simeq20$, with a threshold for the median cluster at $\mu\simeq10$. 

Detailed predictions of the population of multiple images of distant sources formed by the full population of lenses are not feasible at the current time. This is because a model for the covariance of $\thE$, $\etaE$ and $\kE$ across the relevant portion of dark matter halo mass function ($10^{12}\lesssim M_{200}\lesssim10^{15}\,\rm M_\odot$) is available neither from observations nor from simulations. {Constructing this model will be an important step in the future development of this work. Nevertheless, the observed lenses and supporting theoretical work are sufficient to identify the relevant region of parameter space that brackets the expected behaviour of the relevant lenses.} We therefore base our predictions on lens magnification, and exercise caution in interpreting our predictions in terms of multiple imaging in the low-magnification regime ($2<\mu\lesssim10$). We adopt a threshold lens magnification of $\mu>\muth=2$, as this is the lowest magnification at which any of the lenses considered here produce multiple images. 

\subsection{Optical depth}\label{sec:lensing:odsl}

In this context ``optical depth'' refers to the fraction of the celestial sphere that is magnified by a certain amount. It can be expressed either in differential form, $d\tau=\tau_\mu\,d\mu$, as the fraction of the sphere magnified by an amount in the range $\mu\rightarrow\mu+d\mu$, or cumulative form, $\tau(\mu>\muth)$, as the fraction magnified by more than $\muth$. It is also important to distinguish between the optical depth in the image plane, $\tauI$, and the source plane, $\tauS$. The former is the fraction of the observed sky -- the sphere on which the images are located -- that is magnified, whilst the latter is the fraction of the unobservable sphere on which the sources are located that is magnified. In this article we are interested in the number of gravitationally magnified images of a given population of sources. The source-plane optical depth is therefore the most relevant quantity, defined as follows: 
\begin{equation}
  \tau^{\rm S}(\mu>\muth)=\int_{\muth}^{\infty}d\mu\,\frac{\tau^{\rm I}_\mu}{\mu},
  \label{eqn:tauS}
\end{equation}
following \cite{Robertson2020}. In particular, computing $\tauS$ from $\tauI_\mu/\mu$ ensures that both magnification bias is removed and the optical depth is proportional to the number of images that are potentially detectable. Therefore, we predict the number of gravitationally magnified images and thus implicitly integrate over image multiplicity.

Fortunately, the integral of optical depth over lens mass is relatively insensitive to how the structure of gravitational lenses varies with lens mass. Numerous calibrations of $\tau^{\rm S}$ as a function of source redshift agree within a factor of 2 (Figure~\ref{fig:foxetal+odsl}), and show a consistent picture of non-negligible optical depth for dark matter halos spanning masses $10^{12}\,\ls\,M_{200}\,\ls\,10^{15}\,\rm M_\odot$, i.e.\ ranging from individual massive galaxies through to massive galaxy clusters \citep{Robertson2020}. As alluded to in Section~\ref{sec:lensing:versus}, it is also possible to predict the optical depth to multiple imaging by invoking a model for the projected density profile of lenses. We show a recent example of the predicted optical depth to multiple imaging from \cite{Haris2018} as the dashed curve in Figure~\ref{fig:foxetal+odsl}, which they describe with a simple analytic function that we write as:
\begin{equation}
  \tau_{\rm mul}(\zs)=\left[\frac{D^{\rm C}_{\rm S}}{62.2\,{\rm Gpc}}\right]^3.
  \label{eqn:taumul}
\end{equation}
The close agreement between $\tau_{\rm mul}$ and $\tau^{\rm S}(\mu>2)$ in Figure~\ref{fig:foxetal+odsl} is striking, however it masks important subtleties. Predictions of $\tau_{\rm mul}$ assume that the lenses are all individual galaxies with isothermal density profiles \citep[e.g.][]{Haris2018, Oguri2018, Ng2018}, neither of which reflect the true population of lenses as discussed in Section~\ref{sec:lensing:versus}. Moreover, the threshold total magnification at which an SIS lens produces image pairs is $\mup=\mu_++\mu_-=2$. Therefore, formally, \citeauthor{Haris2018}'s $\tau_{\rm mul}=\tau^{\rm S}(\mup>2)$ applies just to SIS lenses. Nevertheless, Equation~\ref{eqn:taumul} is a convenient analytic approximation of $\tau^{\rm S}(\mu\ge\muth,\zs)$, and we therefore use it to derive an analytic expression for the differential source-plane optical depth, $\tau^{\rm S}_\mu$, by setting
\begin{equation}
  \int_{\mu=\muth}^\infty d\mu\,\tau^{\rm S}_\mu(\zs)=\tau^{\rm S}(\mu\ge\muth,\zs)\simeq \tau_{\rm mul}(\zs)
  \label{eqn:taumuth}
\end{equation}
and using the universal scaling $\tau^{\rm S}_\mu\propto\mu^{-3}$ for high magnification \citep{Blandford1986}, to obtain:
\begin{equation}
  \tau^{\rm S}_\mu(\zs)=\left[\frac{D^{\rm C}_{\rm S}}{31.1\,{\rm Gpc}}\right]^3\mu^{-3}.
  \label{eqn:taumus}
\end{equation}
We use Equation~\ref{eqn:taumus} in Section~\ref{sec:gw} when predicting the rate of detectable gravitationally magnified GWs.

\subsection{Time delay}
\label{sec:lensing:time}

\noindent
The arrival time difference between images is directly proportional to the difference between the Fermat potentials (Equation~\ref{eqn:Deltat}). Longer arrival time differences are therefore qualitatively interpreted as being attributable to cluster-scale lenses and shorter differences being attributable to galaxy-scale lenses \citep[e.g.][]{Dai2020}. However, the arrival time difference also depends on the lens magnification and on the structure of the gravitational potential of the lens  \citep[][]{Witt2000,Kochanek2002}. At a deeper level, the arrival time difference for an image pair also depends on the type of lens catastrophe that is responsible for forming the image pair. We discuss this in detail in Appendix~\ref{app:derivation}, where we show that for two common  catastrophes -- the pseudo-caustic of lenses that form just two images (of which the SIS is the archetype) and the fold caustic of lenses that are capable of forming more than two images -- the arrival time difference between a pair of magnified images is given by:
\begin{equation}
    \frac{\Dt_{\rm pseu}}{92\,\rm days}=\left[\frac{\thE}{1''}\right]^2\,\bigg[\frac{\mup}{4}\bigg]^{-1}\,\bigg[\frac{1-\kE}{0.5}\bigg]\,\left[\frac{\mathcal{D}}{3.3\,\rm Gpc}\right],
        \label{eqn:tpseudo}
\end{equation}
\vspace{-4mm}
\begin{equation}
    \frac{\Dt_{\rm fold}}{3.9\,\rm days}=\left[\myfrac[0pt][1pt]{\thE}{1''}\right]^2\left[\myfrac[0pt][1pt]{\mup}{4}\right]^{-3}\left[\myfrac[0pt][1pt]{\etaE}{1}\right]^{-2}\left[\myfrac[0pt][1pt]{\kE}{0.5}\right]^{-2}\left[\myfrac[0pt][1pt]{1-\kE}{0.5}\right]^{-3}\left[\myfrac[0pt][1pt]{\mathcal{D}}{\rm 3.3\,Gpc}\right].
    \label{eqn:tfold}
\end{equation}

\begin{figure}
  \centerline{
    \includegraphics[width=0.8\hsize,angle=0]{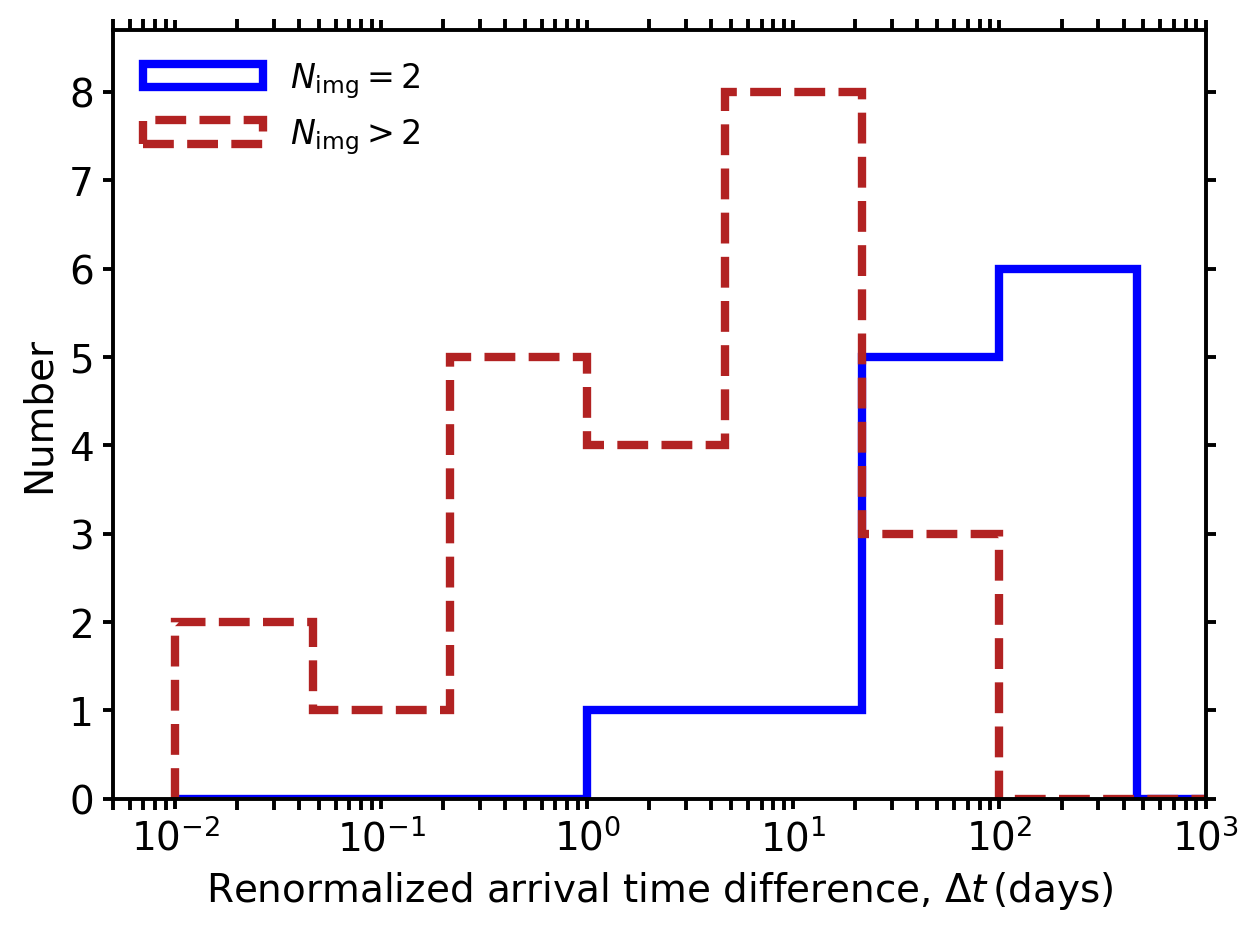}
    }
  \caption{The distribution of arrival time differences obtained by authors cited in Section~\ref{sec:lensing:time} for known time delay lenses including lensed quasars and Supernova Refsdal, split into those for lenses that produced just two images and those that produced more than two images. Arrival time difference has been renormalized to the same lens geometry of $\mD=3.3\,\rm Gpc$ and Einstein radius of $\thE=1\,\rm arcsec$.}\label{fig:dt:verify}
\end{figure}

Equations~\ref{eqn:tpseudo}~\&~\ref{eqn:tfold} are compatible with published measurements of arrival time difference. In Figure~\ref{fig:dt:verify} we show a compilation of arrival time difference measurements from lensed quasars \citep{Fohlmeister2008,Fohlmeister2013,Dahle2015,Millon2020} and supernova Refsdal \citep{Kelly2016,Rodney2016}. The arrival time difference distribution for lenses that produce just two images ($N_{\rm img}=2$) peaks at $\Dt\simeq100\,\rm days$, in good agreement with Equation~\ref{eqn:tpseudo} at low magnification. The tail to smaller $\Dt$ for $N_{\rm img}=2$ is mainly due to a small number of higher magnification pairs. The arrival time difference distribution for lenses that produce more than two images ($N_{\rm img}>2$) is broader and extends to lower $\Dt$, due to the broader range of lensing catastrophes included within this sample, and the stronger dependence of $\Dt$ on $\mu$, $\kE$ and $\etaE$ for fold caustics. For example, these lenses include galaxy-scale quadruply imaged quasars in addition to the wide separation quasar lenses SDSS\,J1004$+$4112, SDSS\,J1029$+$2623, and SDSS\,J2222$+$2745, and the massive galaxy cluster MACS\,J1149.5$+$2223 that was responsible for supernova Refsdal. This distribution extends up to $\Dt\simeq100\,\rm days$, as expected from Equation~\ref{eqn:tpseudo}, and extends down $\Dt\simeq1\,\rm hour$. Such short arrival time differences are compatible with the levels of magnification ($\mup\simeq40$) suffered by image pairs produced by typical galaxy cluster lenses ($\etaE\simeq0.5$, $\kE\simeq0.8$), based on Equation~\ref{eqn:tfold}. 

\section{Gravitationally lensing gravitational waves}\label{sec:gw}

We begin with an overview of the key take away points in Section~\ref{sec:gw:overview}. For readers interested in more detail, we explain how we apply the lensing formalism from Section~\ref{sec:lensing} to GW source populations in Section~\ref{sec:gw:formalism}, which in turn specifies the information that we need about the GW sources. We then describe the populations of NS-NS and BH-BH mergers that we adopt for our lensed GW predictions and verify that they are compatible with the population of sources that have been detected to date in Section~\ref{sec:gw:pops}.

\subsection{Overview}\label{sec:gw:overview}

The basic ingredients required to predict the rates of gravitationally lensed detections of a distant population of objects are (1) the comoving number density of the objects as a function of redshift and a parameter that summarises signal strength, (2) the comoving volume element as function of redshift, which in turn relies on the cosmological model, (3) the lensing probability, and (4) the detector sensitivity. The least well understood of these ingredients will therefore dominate the uncertainties in the predictions. It is therefore immediately apparent that the least certain ingredient is the first, namely the comoving merger rate density of binary compact object mergers, its evolution with redshift, and the mass function of compact objects \citepalias{GWTC3}. In contrast, the cosmological parameters are known to $\simeq10$ per cent, and the optical depth to gravitational magnification is known to a factor of $\simeq2$ (Section~\ref{sec:lensing:odsl}). 

We adopt a range of models for the underlying population of GW sources that are compatible with the population detected to date. This includes mass functions that capture the main features of the BH mass function such as a declining logarithmic slope as a function of mass, possible structure superposed on that broadly declining mass function, and reduced efficiency of BH production from stellar evolution at high mass ($\gtrsim50\,\rm M_\odot$) due to the pair instability. We also include a broad range of evolution models that are compatible with the data, that are variants on the double power law that is commonly used to describe the evolution of the cosmic star formation rate density evolution. 

In Sections~\ref{sec:pred:gw}~\&~\ref{sec:observing} we concentrate on our \emph{Baseline} models of the BH-BH and NS-NS merger populations, as these represent a broad consensus view of the compact object populations. Importantly, in Section~\ref{sec:pred:gw} we establish that departures from our \emph{Baseline} models have only a small effect on the predicted rate of lensed detections. The \emph{Baseline} models and indeed all other models are specified in Table~\ref{tab:pops}. In brief the \emph{Baseline} ties redshift evolution to the evolution of the cosmic star formation rate density, and adopts a BH mass function that spans $5<m<50\,\rm M_\odot$ with a power law slope of $-3.3$, and a flat NS mass function that spans $1<m<2.5\,\rm M_\odot$. We also make several simplifying assumptions that are consistent with our aim to make broad analytic predictions of the lensed population, that are consistent with the data, and that impact our predictions at the level of a factor of $\simeq2$. The most important of these is that we assume equal mass mergers. 

\subsection{Arrival and detection rates}\label{sec:gw:formalism}

The number of gravitationally magnified GWs arriving at Earth per year can be expressed as a function of their apparent mass and distance, $\tM$ and $\tD$, i.e.\ before correction for the effects of lens magnification, as follows:
\begin{equation}
  \frac{d^2\Narr}{d\tM\,d\tD}\,\bigg{|}_{\mu\ge\muth}=
  \int_{z_0}^{\zmax}\,dz\,\frac{\mR(\mM,z)}{1+z}\,\frac{dV}{dz}\,\tau^{\rm S}_\mu(\mu),
  \label{eqn:Narr}
\end{equation}
where $\mR$ is the merger rate per unit comoving volume per unit chirp mass, $dV/dz$ is the comoving volume element, $\mu(z) = [D(z)/\tD]^2$, $z_0$ is the redshift to the distance $D_0$ that satisfies $\mu=\muth=(D_0/\tD)^2$, and as discussed in Section~\ref{sec:lensing:versus} we adopt $\muth=2$. Henceforth, we denote $\zs$ by $z$. The chirp mass, $\mM$, is defined in the usual way:
\begin{equation}
  \mM=\frac{(m_1m_2)^{3/5}}{(m_1+m_2)^{1/5}},
\end{equation}
with all masses scaling with $(1+z)^{-1}$, consequently the mass ratio, $q=m_2/m_1$, is invariant to lens magnification. When needed, we adopt the convention $m_1>m_2$.

The number of gravitationally magnified GWs that are actually detectable per year at Earth depends on the sensitivity of the GW detectors and whether or not they are operating when the signal arrives. We construct $\pdet(\tM,\tD)$ for a given GW run using LIGO (the most sensitive of the GW detectors) sensitivity curves based on \cite{Martynov2016} and take account of the impact of the orientation of the GW detectors with respect to incoming signals from close to the detector's horizon following \cite{Chen2021}. {Although we don't treat the sky localization of GW signals explicitly in this article, we stress that the Virgo and KAGRA detectors play a crucial role in reducing the sky localization uncertainties of GW detections, as for example demonstrated by the detection of GW170817 \citep{GW170817detect}. We return to this in Section~\ref{sec:observing} when we discuss EM follow-up observations.}

We write the total number of magnified GWs that are available to be detected per year at Earth as:
\begin{equation}
  \Ndet =\int_0^{\Dmax}d\tD\int_{\mMmin}^{\mMmax}d\tM\,\,\frac{d^2 \Narr}{d\tM \, d\tD}\,\pdet,
  \label{eqn:ndet}
\end{equation}
with integration limits chosen to span the full region of parameter space over which $\pdet>0$. Note that our approach to describing GW detector sensitivity ignores the presence of a network of detectors and assumes that GW waveforms used to search for GW signals are well matched to the lensed source population. Recent work by \citet{Yang2021} indicates that this contributes a systematic uncertainty of order a factor of 2 to our calculations, which is sub-dominant.

\subsection{Gravitational wave source populations}\label{sec:gw:pops}

We concentrate on NS-NS and BH-BH mergers of roughly equal mass binaries, $q=1$, as these dominate the GW detections to date \citepalias{GWTC3}. The mass of the compact objects that merge are therefore written as $m=m_1=m_2=2^{0.2}\mM$, when convenient to do so below. Note that $\simeq10-20$ per cent departures from $q=1$ have no more that a factor of 2 impact on the volume to which the GW detectors are sensitive, and thus considering only $q=1$ is both well-motivated by data and not a significant source of systematic bias in our calculations. 

We write $\mR$ as a separable function of mass and redshift:
\begin{equation}
  \mR(z,m)=\mR_0\,\,g(z)\,\,f(m)\,\,\varrho^{-1}
  \label{eqn:mRmodel}
\end{equation}
where $\mR_0$ is the comoving merger rate density of binary compact object mergers in the local universe (Section~\ref{sec:gw:local}), $g(z)$ and $f(m)$ are functions that describe the distribution of sources as functions of redshift and compact object mass respectively (Sections~\ref{sec:gw:evolution}~\&~\ref{sec:gw:massfn}). Note that $g(z)$ is normalized to ensure that $g(z=0)=1$ and $f(m)$ is normalized to integrate to unity over the mass range considered. We choose $\varrho$ to ensure the predicted number of GW detections that are not lensed is invariant between the population models (Table~\ref{tab:pops}). 

\subsubsection{Local comoving merger rate density, $\mR_0$}\label{sec:gw:local}

We adopt local comoving merger rate densities based on the credible ranges from \citetalias{GWTC3}, and consistent with \citet{GWTC1,GWTC2}. Formally, these rates assume that the source population does not evolve with redshift, and thus are inconsistent with the broad expectation that the populations do evolve with redshift. However, the difference between local rates based on assuming evolving and non-evolving populations is small compared with other uncertainties \citep{GWTC2}. Moreover, $\varrho\simeq1$ for all models that we consider (Table~\ref{tab:pops}). Therefore, this has negligible impact on our conclusions, and we adopt $\varrho=1$ for all models, $13<\mR_0<1900\,\rm Gpc^{-3}yr^{-1}$ for NS-NS, and $16<\mR_0<130\,\rm Gpc^{-3}yr^{-1}$ for BH-BH.

\subsubsection{Redshift evolution, $g(z)$}
\label{sec:gw:evolution}

\begin{table*}
  \caption{Models for gravitational wave source populations}
  \label{tab:pops}
  \centering
  \begin{tabular}{lcccccccccccccccc}
    \hline
    Model        & ~~ & \multispan{2}{Integration limits} & ~~ & \multispan{6}{Mass function\dotfill} & ~~ & \multispan{3}{Evolution\dotfill} & ~~ & $\varrho$ \cr
                 &    & $\mmin$        & $\mmax$          &    & $\gamma_1$ & $\gamma_2$ & $\mbr$         & $m_0$          & $\sigmam$     & $\phip$ & & $\alpha$ & $\beta$ & $\zp$ & & \cr
                 &    & $\rm(M_\odot)$ & $\rm(M_\odot)$   &    &            &            & $\rm(M_\odot)$ & $\rm(M_\odot)$ & $\rm(M_\odot)$ & \cr
    \hline
    \noalign{\smallskip}
    \multispan{5}{\underline{NS-NS mergers}}\cr
    \noalign{\smallskip}
    Baseline     & & $1$ & $2.5$ & & $0$ & $0$ & ... & ... & ...& $0$ & & $2.7$ & $2.9$ & $1.9$ & & $1.00$ \cr
    \noalign{\smallskip}
    Steep        & & $1$ & $2.5$ & & $-2$& $-2$ & ... & ... & ... & $0$ & & $2.7$ & $2.9$ & $1.9$ & & $0.73$ \cr
    Light        & & $1$ & $2$   & & $0$ & $0$ & ... & ... & ... & $0$ & & $2.7$ & $2.9$ & $1.9$ & & $0.65$ \cr
    Heavy        & & $1$ & $3$   & & $0$ & $0$ & ... & ... & ... & $0$ & & $2.7$ & $2.9$ & $1.9$ & & $1.45$ \cr
    \noalign{\smallskip}
    Delayed  & & $1$ & $2.5$ & & $-2$ & $-2$ & ... & ... & ... & $0$ & & $3.3$ & $2.7$ & $1.5$ & & $1.03$ \cr
    Maximal  & & $1$ & $2.5$ & & $-2$ & $-2$ & ... & ... & ... & $0$ & & $4.5$ & $2.5$ & $1.0$ & & $1.08$ \cr
    Power-law & & $1$ & $2.5$ & & $-2$ & $-2$ & ... & ... & ... & $0$ & & $2.7$ & $0$ & $\infty$ & & $1.00$ \cr
    Non-evolving & & $1$ & $2.5$ & & $-2$ & $-2$ & ... & ... & ... & $0$ & & $0$ & $0$ & $\infty$ & & $0.89$ \cr
    \noalign{\smallskip}
    \hline
    \noalign{\smallskip}
    \multispan{5}{\underline{BH-BH mergers}}\cr
    \noalign{\smallskip}
    Baseline  & & $5$ & $500$ & & $-3.3$ & $-\infty$ & $50$ & ... & ... & $0$ & & $2.7$ & $2.9$ & $1.9$ & & $1.00$ \cr
    \noalign{\smallskip}
    Tapered   & & $5$ & $500$ & & $-3.3$ & $-6$  & $50$ & ...  & ... & $0$ &  & $2.7$ & $2.9$ & $1.9$ & & $1.28$ \cr
    Steep     & & $5$ & $500$ & & $-3.3$ & $-16$ & $50$ & ...  & ... & $0$ &  & $2.7$ & $2.9$ & $1.9$ & & $1.21$ \cr
    Perturbed & & $5$ & $500$ & & $-3.3$ & $-16$ & $50$ & $33$ & $2$ & $0.03$ & & $2.7$ & $2.9$ & $1.9$ & & $1.81$ \cr
    \noalign{\smallskip}
    Delayed  & & $5$ & $500$ & & $-3.3$ & $-\infty$ & $50$ & ... & ... & $0$ & & $3.3$ & $2.7$ & $1.5$ & & $1.36$ \cr
    Maximal  & & $5$ & $500$ & & $-3.3$ & $-\infty$ & $50$ & ... & ... & $0$ & & $4.5$ & $2.5$ & $1.0$ & & $1.79$ \cr
    Power-law & & $5$ & $500$ & & $-3.3$ & $-16$ & $50$ & ... & ... & $0$ & & $2.7$ & $0$ & $\infty$ & & $1.19$ \cr
    Non-evolving & & $5$ & $500$ & & $-3.3$ & $-16$ & $50$ & ... & ... & $0$ & & $0$ & $0$ & $\infty$ & & $0.64$ \cr
    \noalign{\smallskip}
    \hline 
  \end{tabular}
\end{table*}

We parameterize redshift evolution as a double power law for both NS-NS and BH-BH mergers:
\begin{equation}
  g(z)=C(1+z)^\alpha\left[1+\left(\frac{1+z}{1+\zp}\right)^{\alpha+\beta}\right]^{-1}
  \label{eqn:madau}
\end{equation}
where $\zp$ is the redshift at which the redshift evolution pivots from $g\propto(1+z)^\alpha$ to $g\propto(1+z)^{-\beta}$, and $C=1+(1+\zp)^{-(\alpha+\beta)}$ ensures that $g(z=0)=1$, following \cite{Callister2020}. 

Our \emph{Baseline} model adopts the canonical values of $\alpha=2.7$, $\beta=2.9$, and $\zp=1.9$, based on measurements of the star formation rate density (SFRD) history of the universe \citep[e.g.][and references therein]{Madau2014}. This is equivalent to assuming that the elapsed time from star formation to merger of binary compact objects is zero. Whilst this is clearly unphysical, connecting $g(z)$ with the cosmic SFRD is a physically well motivated ansatz, and broadly consistent with population synthesis models of binary compact object mergers \citep[e.g.][]{Dominik2013,Santoliquido2021}. This model is consistent with the GW source populations detected to date \citepalias{GWTC3}. {Moreover, all but the most extreme delay times between star formation and binary compact object mergers -- approaching the age of the universe -- have a negligible effect on the predicted rate of lensed GWs \citep{Mukherjee2021delay}. We explore and confirm this by experimenting with alternative models below.}

Whilst \citetalias{GWTC3}'s detection of redshift evolution is strong, their constraints on $\alpha$ (their $\kappa$) are quite uncertain ($\alpha=2.7^{+1.8}_{-1.9}$), and only extend to $z\simeq1$, i.e.\ $2.4\,\rm Gyr$ later than the pivot redshift $\zp=1.9$.  We therefore experiment with a \emph{Delayed} model ($\alpha=3.3$, $\beta=2.7$, $\zp=1.5$) and a \emph{Maximal} model ($\alpha=4.5$, $\beta=2.5$, $\zp=1.0$). The parameters of these models were chosen by fixing the pivot redshift at $\zp=1.5$ and $\zp=1$ respectively so as to bracket the range of $\zp$ compatible with the cosmic star formation rate history and GW constraints on evolution to date. This forces $\alpha$ for these two models to be steeper than for the \emph{Baseline} model. We chose $\alpha=4.5$ for the \emph{Maximal} model, i.e.\ at the steep end of \citetalias{GWTC3}'s constraints, then obtained $\beta=2.5$ for this model by requiring the integral of $g(z)$ to be invariant to choice of model parameters -- i.e. invariance of the total number of stellar remnant compact object binaries in the universe. We followed a similar procedure to choose $\alpha$ and $\beta$ for the \emph{Delayed} model. For reference, we also use \emph{Non-evolving} ($\alpha=0$, $\beta=0$, $\zp\rightarrow\infty$) and \emph{Power-law} ($\alpha=2.7$, $\beta=0$, $\zp\rightarrow\infty$) models.

\subsubsection{Mass function, $f(m)$}\label{sec:gw:massfn}

\begin{figure}
  \centerline{
    \includegraphics[width=\hsize,angle=0]{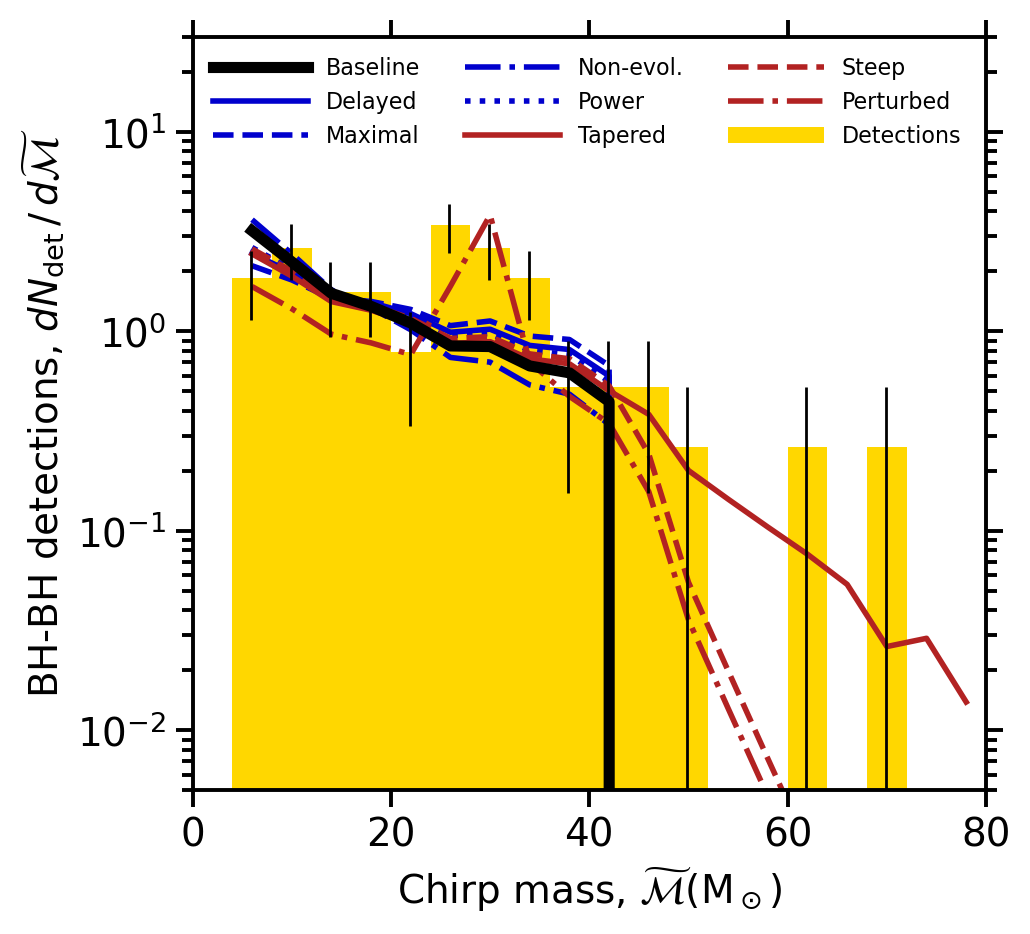}
  }
  \caption{The source-plane chirp mass distribution of objects with $\widetilde{\mathcal{M}}>5/2^{0.2}\simeq4\,\rm M_\odot$ detected by LIGO and Virgo in their third run, compared with that computed from our BH mass function models (Section~\ref{sec:gw:massfn}), using Equation~\ref{eqn:ndet} with lensing turned off. The model curves adopt $\mR_0=18.3\,\rm Gpc^{-3}\,yr^{-1}=30/1.2^{2.7}$ based on the ``best constrained'' BH-BH merger rate at $z=0.2$ given by \citetalias{GWTC3}, and reflect a run duration of 11 months with detectors in operation 70 per cent of the time.}\label{fig:mchirpdist}
\end{figure}

In general, we parameterise the mass function as a double power law:
\begin{equation}
  f(m)\propto
  \begin{cases}
    ~0&~:~m<\mmin\\
    ~(m/m_{\rm br})^{\gamma_1}&~:~\mmin<m<\mbr\\
    ~(m/m_{\rm br})^{\gamma_2}&~:~\mbr<m<\mmax\\
    ~0&~:~m>\mmax,\\
  \end{cases}
\end{equation}
where $\mbr$ is the ``break'' at which the power law slope changes, and $\mmin$ and $\mmax$ are effectively the integration limits of the mass function. 

For NS-NS mergers we adopt a \emph{Baseline} model with $\gamma_1=\gamma_2=0$, $\mmin=1\,\rm M_\odot$, and $\mmax=2.5\,\rm M_\odot$, in line with \citetalias{GWTC3}. To test the sensitivity of the predicted rate of lensed NS-NS mergers to these choices, we also use a \emph{Heavy} model that extends to $\mmax=3\,\rm M_\odot$, a \emph{Light} model that is restricted to $\mmax=2\,\rm M_\odot$, and a \emph{Steep} model with $\gamma_1=\gamma_2=-2$.

For BH-BH mergers we explore a range of mass functions that are  consistent with the theoretical expectation that the efficiency of BH production by stellar evolution declines at $m\gs50\,\rm M_\odot$ due to the pair instability, and the broad features of the GW population detected to date, namely a steeply declining BH mass function and possible structure (i.e.\ peaks) in the mass function \citepalias{GWTC3}. Our \emph{Baseline} BH mass function is an unbroken power law, implemented as $\gamma_1=-3.3$, $\gamma_2\rightarrow-\infty$, $\mbr=50\,\rm M_\odot$, and with $5<m<500\,\rm M_\odot$. This model effectively mandates that detections of BH-BH mergers consistent with component masses $m>\mbr=50\,\rm M_\odot$ are gravitationally lensed -- a physically plausible model that is based on well understood physics. However, $\gamma_2\rightarrow-\infty$ may be rather extreme, and so our \emph{Tapered} model differs from \emph{Baseline} by a less aggressive reduction in efficiency of BH production via $\gamma_2=-6$, and is qualitatively able to match the detection of a few very massive BH-BH mergers without mandating that they are lensed. Our \emph{Steep} model is intermediate between \emph{Baseline} and \emph{Tapered}, with $\gamma_2=-16$, which has been tuned to give a comparable number of apparently very massive BH-BH mergers from both lensing and massive BH formation channels. Finally, our \emph{Perturbed} model differs from our \emph{Steep} model by adding a normally distributed perturbation that is centred at $m_0=33\,\rm M_\odot$, with width $\sigmam=2\,\rm M_\odot$, and a fraction of the BHs described by the mass function that inhabit this perturbation of $\phip=0.03$. 

As a consistency check, we show in Figure~\ref{fig:mchirpdist} that all of the BH models are able to reproduce the broad features of the source-frame chirp mass distribution of BH-BH mergers from LIGO and Virgo's third run. We use the population of detected BH-BH mergers for this purpose, and not NS-NS merger detections, because of the small numbers of the latter detected to date. 

\section{Rates and properties of gravitationally lensed gravitational waves}
\label{sec:pred:gw}

We present our predictions for the population of gravitationally magnified GWs from BH-BH and NS-NS mergers, concentrating on the \emph{Baseline} model described in Section~\ref{sec:gw:pops} and summarised in Table~\ref{tab:pops}, discussing the stability of our predictions to model choice where relevant. The predictions are organised as follows: detection rates and discovery timescales (Section~\ref{sec:pred:gw:rates}), distance and magnification distributions (Section~\ref{sec:pred:gw:dmudist}), mass distributions (Section~\ref{sec:pred:gw:mdist}), and time delay distributions (Section~\ref{sec:pred:gw:dtdist}). We compare our preductions with previous work in Section~\ref{sec:pred:compare} and summarise the implications of our predictions for discovery strategies in Section~\ref{sec:pred:implications}.

\subsection{Rates of detectable lensed BH-BH and NS-NS mergers}
\label{sec:pred:gw:rates}

We predict that in the fifth GW run $\simeq0.1$ per cent of NS-NS merger detections and $\simeq0.2$ per cent of BH-BH merger detections will be magnified by $\mu>2$ (Table~\ref{tab:relative}). These relative rates are $\simeq5\times$ and $\simeq2\times$ larger than in the first run for NS-NS and BH-BH respectively. Interestingly, the relative rates for lensed NS-NS mergers in the fifth GW run are comparable with the relative rates for lensed BH-BH mergers in the first run. Our predictions are stable to varying the source population models, with the  relative rate of lensed detections varying by just $\simeq10-20$ per cent between models (Table~\ref{tab:relative}). The only exceptions are the \emph{Non-evolving} and \emph{Power-law} models, that have been included here for reference. 

\begin{table}
  \caption{Predicted relative rates of lensed GW detections as a function of GW detector sensitivity and compact object mass function, expressed as the ratio of the number of detectable events magnified by $\mu>2$ to the number of detectable events that are not lensed.}
  \begin{tabular}{llllll}
    \hline
    Model & O1 & O2 & O3 & O4 & O5\cr
    &    &    &    & ``Design'' & ``A$+$'' \cr
    \hline
    \noalign{\smallskip}
    \multispan{3}{\underline{NS-NS mergers:}}\cr
    \noalign{\smallskip}
    Baseline & $1{:}4400$	& $1{:}3500$	& $1{:}2500$	& $1{:}1900$	& $1{:}970$\cr
    Heavy & $1{:}3900$	& $1{:}3100$	& $1{:}2200$	& $1{:}1700$	& $1{:}860$\cr
    Light & $1{:}5200$	& $1{:}4100$	& $1{:}2900$	& $1{:}2300$	& $1{:}1100$\cr
    Steep & $1{:}4800$	& $1{:}3800$	& $1{:}2700$	& $1{:}2100$	& $1{:}1000$\cr
    Delayed & $1{:}3900$	& $1{:}3000$	& $1{:}2200$	& $1{:}1700$	& $1{:}890$\cr
    Maximal & $1{:}3500$	& $1{:}2800$	& $1{:}2000$	& $1{:}1600$	& $1{:}890$\cr
    Non-evolving & $1{:}21000$	& $1{:}16000$	& $1{:}11000$	& $1{:}8600$	& $1{:}3800$\cr
    Power-law & $1{:}660$	& $1{:}510$	& $1{:}360$	& $1{:}270$	& $1{:}130$\cr
    \hline
    \noalign{\smallskip}
    \multispan{3}{\underline{BH-BH mergers:}}\cr
    \noalign{\smallskip}
    Baseline & $1{:}1100$	& $1{:}870$	& $1{:}750$	& $1{:}570$	& $1{:}520$\cr
    Tapered & $1{:}1100$	& $1{:}880$	& $1{:}770$	& $1{:}590$	& $1{:}550$\cr
    Steep & $1{:}1100$	& $1{:}850$	& $1{:}750$	& $1{:}570$	& $1{:}540$\cr
    Perturbed & $1{:}1000$	& $1{:}800$	& $1{:}740$	& $1{:}540$	& $1{:}600$\cr
    Delayed & $1{:}980$	& $1{:}790$	& $1{:}710$	& $1{:}580$	& $1{:}600$\cr
    Maximal & $1{:}980$	& $1{:}840$	& $1{:}810$	& $1{:}730$	& $1{:}820$\cr
    Non-evolving & $1{:}4100$	& $1{:}3000$	& $1{:}2400$	& $1{:}1600$	& $1{:}1100$\cr
    Power-law & $1{:}320$	& $1{:}230$	& $1{:}210$	& $1{:}120$	& $1{:}100$\cr
    \hline
  \end{tabular}
  \label{tab:relative}
\end{table}

\begin{table*}
  \caption{Predicted absolute number of detectable lensed (magnified
    by $\mu\ge2$) GWs per year as a function of GW detector sensitivity and compact object mass function. }
  \begin{tabular}{llllll}
    \hline
    & O1 & O2 & O3 & O4 (Design) & O5 (A$+$)\cr
    \hline
    \noalign{\smallskip}
    \hspace{-1.8mm}\underline{NS-NS mergers:}\cr
    \noalign{\smallskip}
Baseline & $1\times10^{-5}-2\times10^{-3}$ & $3\times10^{-5}-5\times10^{-3}$ & $1\times10^{-4}-2\times10^{-2}$ & $3\times10^{-4}-5\times10^{-2}$ & $7\times10^{-3}-1.0$\\
Heavy & $2\times10^{-5}-3\times10^{-3}$ & $5\times10^{-5}-8\times10^{-3}$ & $2\times10^{-4}-3\times10^{-2}$ & $6\times10^{-4}-8\times10^{-2}$ & $1\times10^{-2}-1.7$\cr
Light & $7\times10^{-6}-1\times10^{-3}$ & $2\times10^{-5}-3\times10^{-3}$ & $7\times10^{-5}-1\times10^{-2}$ & $2\times10^{-4}-3\times10^{-2}$ & $4\times10^{-3}-0.6$\cr
Steep & $8\times10^{-6}-1\times10^{-3}$ & $2\times10^{-5}-3\times10^{-3}$ & $9\times10^{-5}-1\times10^{-2}$ & $2\times10^{-4}-3\times10^{-2}$ & $5\times10^{-3}-0.7$\cr
Delayed & $1\times10^{-5}-2\times10^{-3}$ & $4\times10^{-5}-5\times10^{-3}$ & $2\times10^{-4}-2\times10^{-2}$ & $4\times10^{-4}-6\times10^{-2}$ & $8\times10^{-3}-1.2$\cr
Maximal & $2\times10^{-5}-2\times10^{-3}$ & $4\times10^{-5}-6\times10^{-3}$ & $2\times10^{-4}-2\times10^{-2}$ & $5\times10^{-4}-7\times10^{-2}$ & $9\times10^{-3}-1.3$\cr
Non-evolving & $2\times10^{-6}-3\times10^{-4}$ & $6\times10^{-6}-9\times10^{-4}$ & $3\times10^{-5}-4\times10^{-3}$ & $7\times10^{-5}-1\times10^{-2}$ & $1\times10^{-3}-0.2$\cr
Power-law & $8\times10^{-5}-1\times10^{-3}$ & $2\times10^{-4}-3\times10^{-2}$ & $9\times10^{-4}-0.1$ & $2\times10^{-3}-0.4$ & $5\times10^{-2}-7.4$\cr
\hline
    \noalign{\smallskip}
    \hspace{-1.8mm}\underline{BH-BH mergers:}\cr
    \noalign{\smallskip}
Baseline & $4\times10^{-3}-4\times10^{-2}$ & $1\times10^{-2}-0.1$ & $4\times10^{-2}-0.3$ & $0.1-1.0$ & $1.4-11$\cr
Tapered & $6\times10^{-3}-5\times10^{-2}$ & $2\times10^{-2}-0.1$ & $5\times10^{-2}-0.4$ & $0.2-1.3$ & $1.6-13$\cr
Steep & $5\times10^{-3}-4\times10^{-2}$ & $2\times10^{-2}-0.1$ & $5\times10^{-2}-0.4$ & $0.2-1.3$ & $1.6-13$\cr
Perturbed & $9\times10^{-3}-7\times10^{-2}$ & $2\times10^{-2}-0.2$ & $8\times10^{-2}-0.6$ & $0.3-2.0$ & $1.9-16$\cr
Delayed & $6\times10^{-3}-5\times10^{-2}$ & $2\times10^{-2}-0.1$ & $6\times10^{-2}-0.5$ & $0.2-1.4$ & $1.8-14$\cr
Maximal & $8\times10^{-3}-6\times10^{-2}$ & $2\times10^{-2}-0.2$ & $7\times10^{-2}-0.6$ & $0.2-1.6$ & $1.8-15$\cr
Non-evolving & $1\times10^{-3}-8\times10^{-3}$ & $3\times10^{-3}-2\times10^{-2}$ & $8\times10^{-3}-7\times10^{-2}$ & $3\times10^{-2}-0.2$ & $0.2-2$\cr
Power-law & $2\times10^{-2}-0.2$ & $6\times10^{-2}-0.4$ & $0.2-1.5$ & $0.7-6.0$ & $9.0-73$\cr
\hline    
  \end{tabular}
  \label{tab:absolute}
\end{table*}

We predict that the number of detectable gravitationally lensed BH-BH mergers will for the first time approach one per year in the fourth GW run (Table~\ref{tab:absolute}), with a handful of detections per year in the fifth run. We also predict that the number of detectable gravitationally lensed NS-NS mergers will lag just one run behind lensed BH-BH mergers, with up to $\simeq1$ lensed NS-NS being detectable per year during the fifth run (Table~\ref{tab:absolute}). The dominant uncertainty on these rates is the uncertainty on the local rates of compact object mergers \citepalias{GWTC3}. Importantly, the predicted rates vary by less than a factor two across the relevant source population models, with only the \emph{Non-evolving} and \emph{Power-law} models deviating strongly.

\begin{figure}
  \centerline{
    \includegraphics[width=\hsize,angle=0]{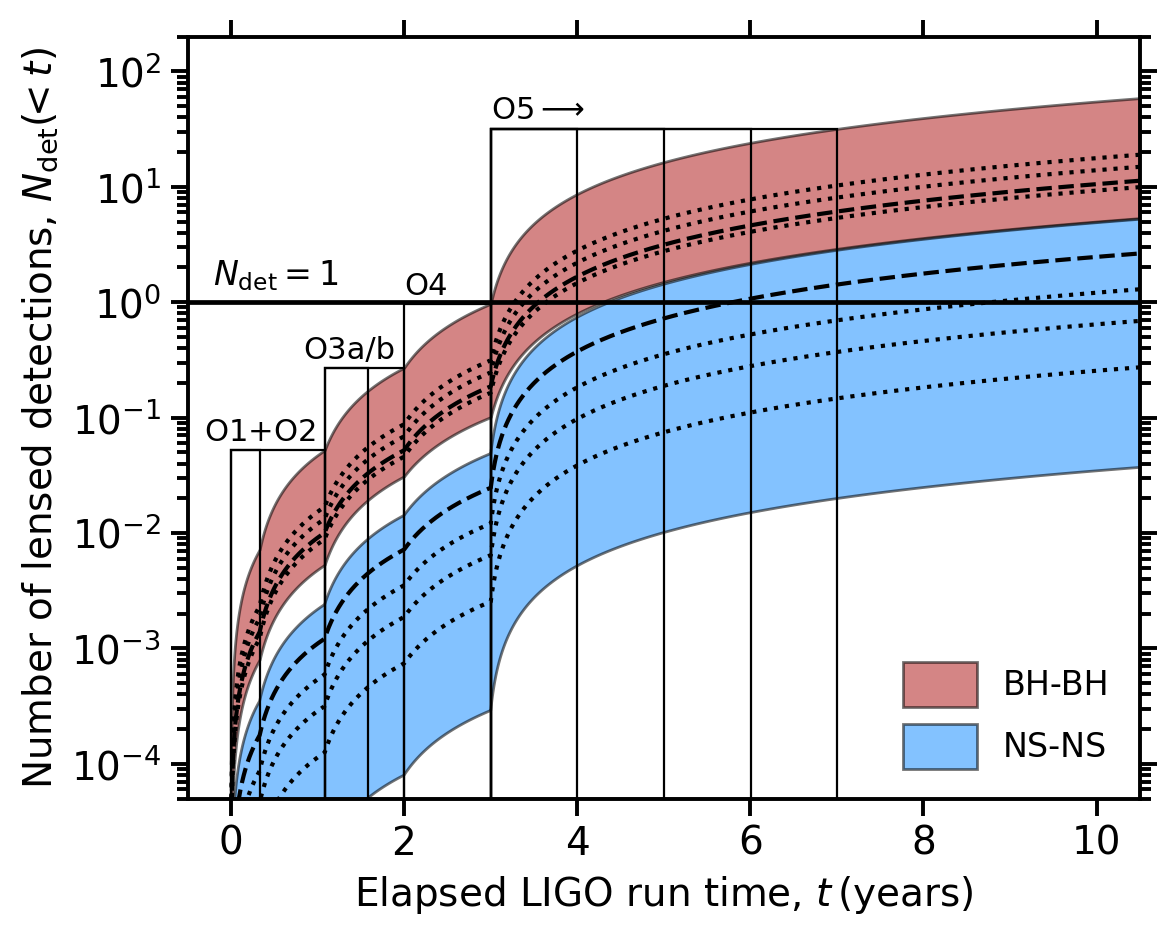}
  }
  \caption{Predicted cumulative number of detected gravitationally magnified BH-BH (red upper shaded band) and NS-NS (blue lower shaded band) mergers for the \emph{Baseline} models described in Section~\ref{sec:gw:pops} and Table~\ref{tab:pops}. Detection of {magnified} BH-BH mergers are predicted to be a regular occurrence in the fifth GW run and beyond, and detection of {magnified} NS-NS mergers are predicted to become more likely the longer that LIGO in particular operates at its A$+$ sensitivity or better in the fifth run. In this figure $\Ndet$ is lower than the integral of $\Ndet$ from Equation~\ref{eqn:ndet} by a factor equal to the GW detector duty cycle, which is taken to be $0.6$ in the first two runs and $0.7$ in later runs. The durations of the first three runs are matched to their actual duration, the fourth run is assumed to last one year (expected to {start in} 2023), and the fifth and later runs are assumed to span multiple years in to the late 2020s. The horizontal solid line marks the detection of one {magnified} source, $N_{\rm det}=1$. The width of the shaded bands matches the ranges quoted on the values listed in Table~\ref{tab:absolute}, i.e.\ they reflect the full range of uncertainties discussed by \citetalias{GWTC3}. The dotted curves show predictions that correspond to the central values of the three compact object mass functions from \citetalias{GWTC3} that contribute to the ranges on $\mR_0$ that we adopt; the dashed curves correspond to \citetalias{GWTC3}'s ``PDB (pair)'' model. }\label{fig:towards}
\end{figure}

We combine the predicted detection rates from Table~\ref{tab:absolute} with the global fraction of an LVK run for which the GW detectors are obtaining data to show the predicted cumulative number of detectable GWs as a function of run time in Figure~\ref{fig:towards}. The predicted cumulative number of gravitationally magnified compact object merger detections grows steadily with time, with detection becoming routine in the mid-2020s. Given the strong prospects for first detection in upcoming runs, it is important to consider how to recognise a lensed detection among the growing number of detections that will be made. We address this in the following Sections. 

\subsection{Distance and magnification distributions}\label{sec:pred:gw:dmudist}

\begin{figure}
  \centerline{
    \includegraphics[width=0.8\hsize,angle=0]{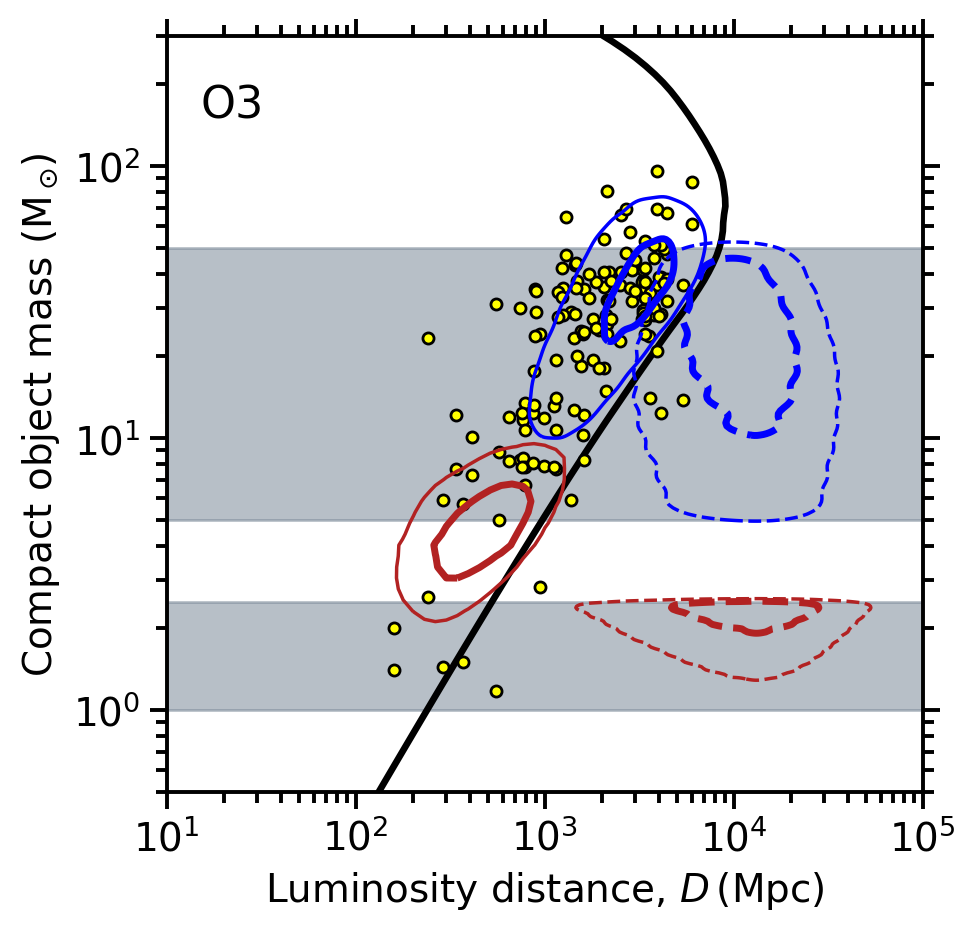}
  }
  \vspace{2mm}
  \centerline{
    \includegraphics[width=0.8\hsize,angle=0]{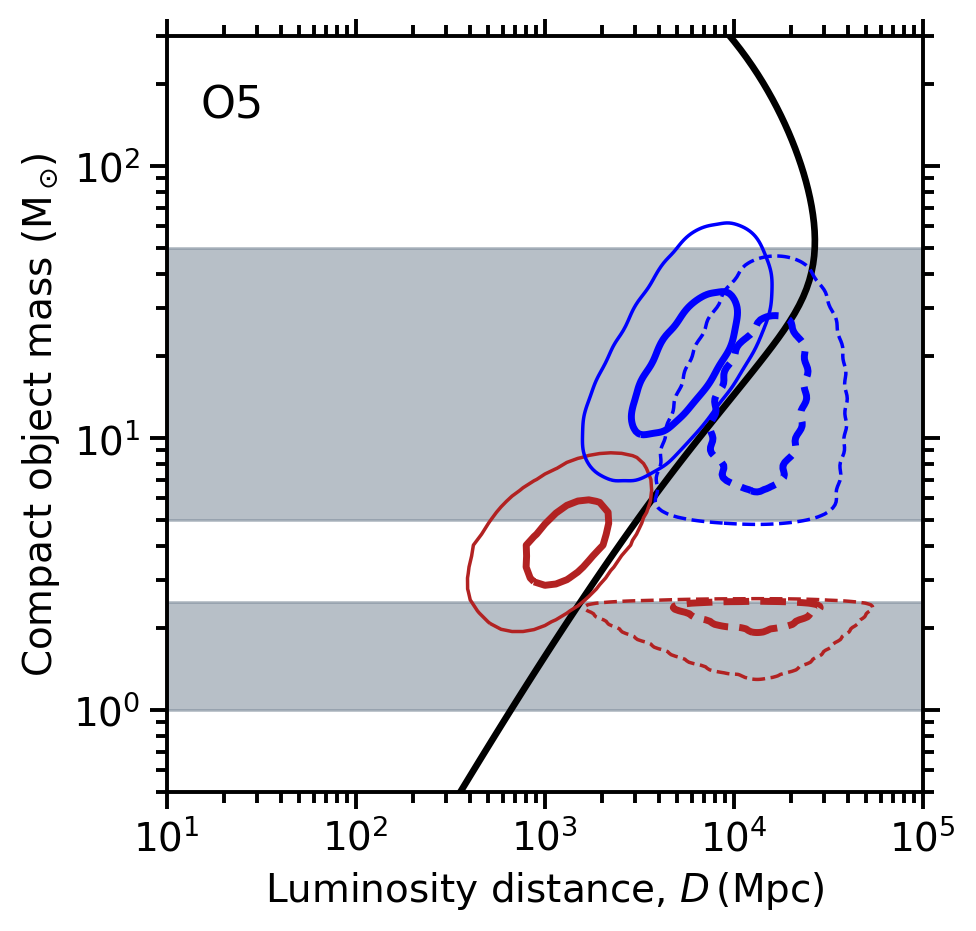}
  }
  \caption{Predicted distributions of gravitationally magnified NS-NS (red contours) and BH-BH (blue contours) merger detections in the third (top panel) and fifth (bottom panel) GW runs for our \emph{Baseline} models. Solid and dashed contours show the $\tm$-$\tD$ (i.e.\ as inferred in low latency by the LVK collaboration assuming $\mu=1$) and $m$-$D$ (i.e.\ true intrinsic values) distributions respectively. In each case the thicker (inner) and thinner (outer) contours encircle $50$ and $90$ per cent of the predicted magnified population respectively. Grey horizontal bands show the mass range encompassed by the mass functions in our \emph{Baseline} models. The points show the masses of the individual compact objects that comprise the population of binary compact object mergers detected through to the end of the third GW run. The thick black curve in each panel shows LIGO's horizon for equal mass mergers based on \citet{Martynov2016}.}
  \label{fig:contours}
\end{figure}

\begin{figure}
  \centerline{
  \includegraphics[width=0.8\hsize,angle=0]{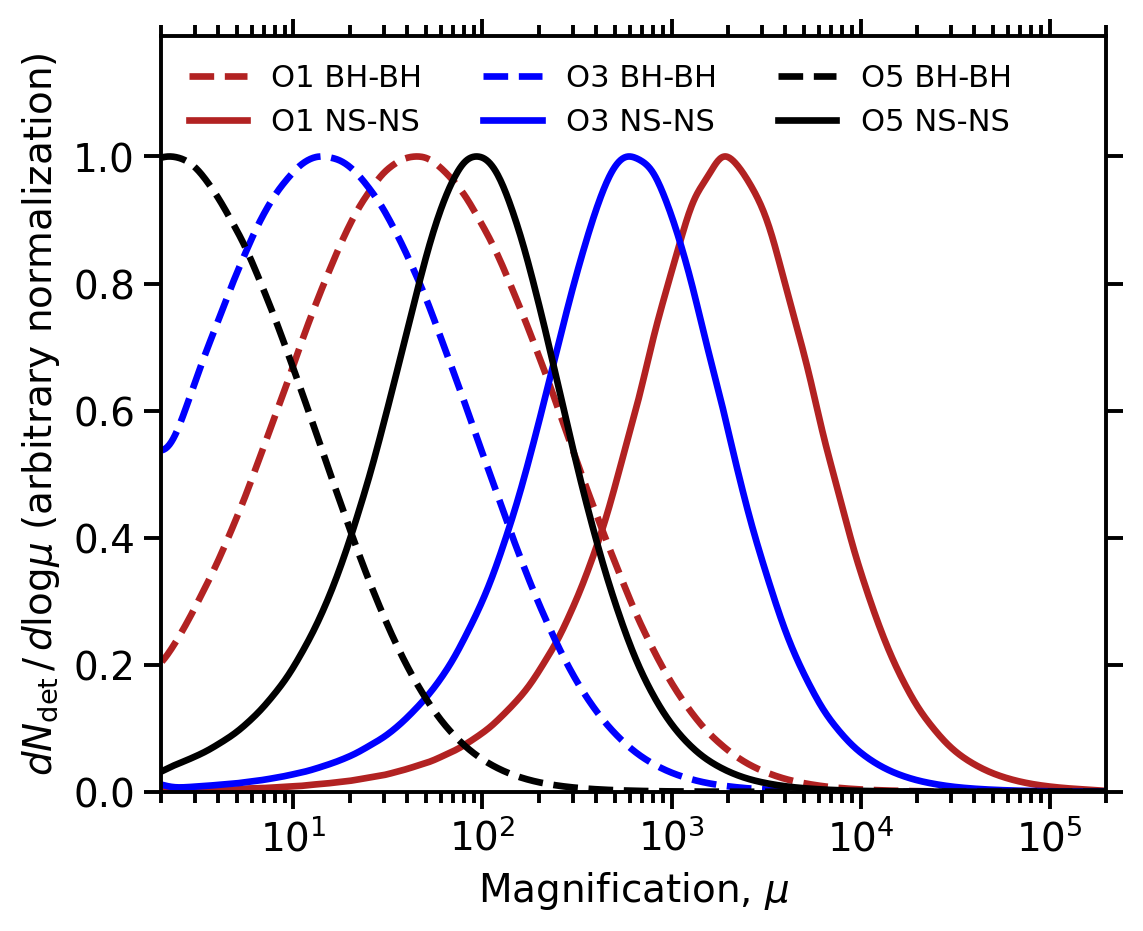}
  }
  \caption{{Predicted lens magnification distributions as a function of sensitivity achieved/forecast in the first, third and fifth GW run, for lensed BH-BH (dashed) and lensed NS-NS (solid) mergers, and based on our \emph{Baseline} models. The black (left-most) dashed and solid curves show that in the fifth run lensed BH-BH will be dominated by low-magnification events $\mu\simeq2-10$, whilst lensed NS-NS will typically magnified by $\mu\simeq100$. The latter is comparable with the typical lens magnifications suffered by putative lensed BH-BH in the first run (right-most dashed curve) in the first run. The predicted rate of detectable lensed NS-NS in O5 is higher than for detectable lensed BH-BH in O1 (Table~\ref{tab:absolute}) because the intrinsic comoving merger rate density of mergers is higher for NS-NS than for BH-BH \citepalias{GWTC3}. In general, the curves move to the left (lower lens magnification required for detection) as the GW detectors become more sensitive, as seen for example in the LIGO horizon in O5 being rightward of their horizon in O3 in Figure~\ref{fig:contours}.}}
  \label{fig:mudist}
\end{figure}

Gravitationally lensed BH-BH and NS-NS mergers are detected as being louder and closer to Earth than they really are, as a consequence of the lens magnification (Equation~\ref{eqn:mubias}). We predict that the centre of the true mass-distance ($m$-$D$) distribution of both lensed BH-BH and lensed NS-NS mergers is located at $D\simeq15\,\rm Gpc$ for all five GW runs considered in this article (dashed contours in Figure~\ref{fig:contours}). As the GW detectors become more sensitive from one run to the next, the distance out to which it is possible to detect gravitationally magnified GWs increases, as seen in the rightward shift of the solid contours between left and central panels in Figure~\ref{fig:contours}. This signifies the gradual reduction in the lens magnification required to bring faint distant sources within the LIGO horizon. This is summarised in Figure~\ref{fig:mudist}, where we see that the typical lens magnification relevant to BH-BH mergers reduces from $\mu\simeq50$ in the first run to $\mu\simeq2$ in the fifth run, and relevant to NS-NS mergers from $\mu\simeq2000$ in the first run to $\mu\simeq100$ in the fifth run. {Therefore in the fifth run, the lens population responsible for detectable multiply-imaged BH-BH mergers will be dominated by galaxy-scale lenses, with cluster-scale lenses also contributing detectable singly-imaged magnified BH-BH mergers. In contrast, lenses of all mass scales will contribute to the detectable multiply-imaged NS-NS population in the fifth run, with a negligible fraction of detectable singly-imaged NS-NS mergers located behind cluster-scale lenses.} 

\subsection{Mass distributions}\label{sec:pred:gw:mdist}

In this Section we use Figure~\ref{fig:contours} to consider the $\tm$ distribution as a tool for selecting candidate lensed GWs. 

The majority of lensed BH-BH mergers do not have anomalously large $\tm$, with just 20 per cent and 5 per cent of them having $\tm>50\,\rm M_\odot$ in LIGO/Virgo's third and fifth runs respectively. The ``ordinariness'' of the predicted $\tm$ values of lensed BH-BH mergers is caused by the declining slope of the BH mass function \citepalias{GWTC3}, as implemented in all of our mass function models (Table~\ref{tab:pops}) -- i.e.\ there are relatively few BH-BH mergers of intrinsically high mass available to be magnified so as to appear to have anomalously large mass. Selecting candidate lensed BH-BH mergers based on $\tm$ is therefore inefficient. We stress that we do not interpret Figure~\ref{fig:contours} as indicating that most BH-BH detections to date are lensed, and draw attention to the relative detection rates of lensed GWs listed in Table~\ref{tab:relative}.

Selecting candidate lensed NS-NS based on $\tm$ appears promising because the predicted $\tm$-distribution of lensed NS-NS mergers peaks within the mass gap between NS and BH (Figures~\ref{fig:contours}). Crucially, the lensed NS-NS $\tm$ distribution \emph{peaks} in this region of parameter space, in contrast to the \emph{tail} of lensed BH-BH mergers at $\tm>50\,\rm M_\odot$ discussed above. The fraction of lensed NS-NS mergers predicted to be inferred by the LVK collaboration to be in this mass gap at $2.5<\tm<5\,\rm M_\odot$ is 62 per cent in their fifth run -- i.e. $\simeq12\times$ more efficient as an approach for candidate lensed GW selection than the 5 per cent discussed above for lensed BH-BH candidates. This is insensitive to the choice of NS population model (Table~\ref{tab:pops}): $\gtrsim50$ per cent of lensed NS-NS detections reside in the lower mass gap for all of the NS mass functions considered.

\subsection{Time delay distributions}\label{sec:pred:gw:dtdist}

\begin{figure*}
  \centerline{
    \includegraphics[width=0.7\hsize,angle=0]{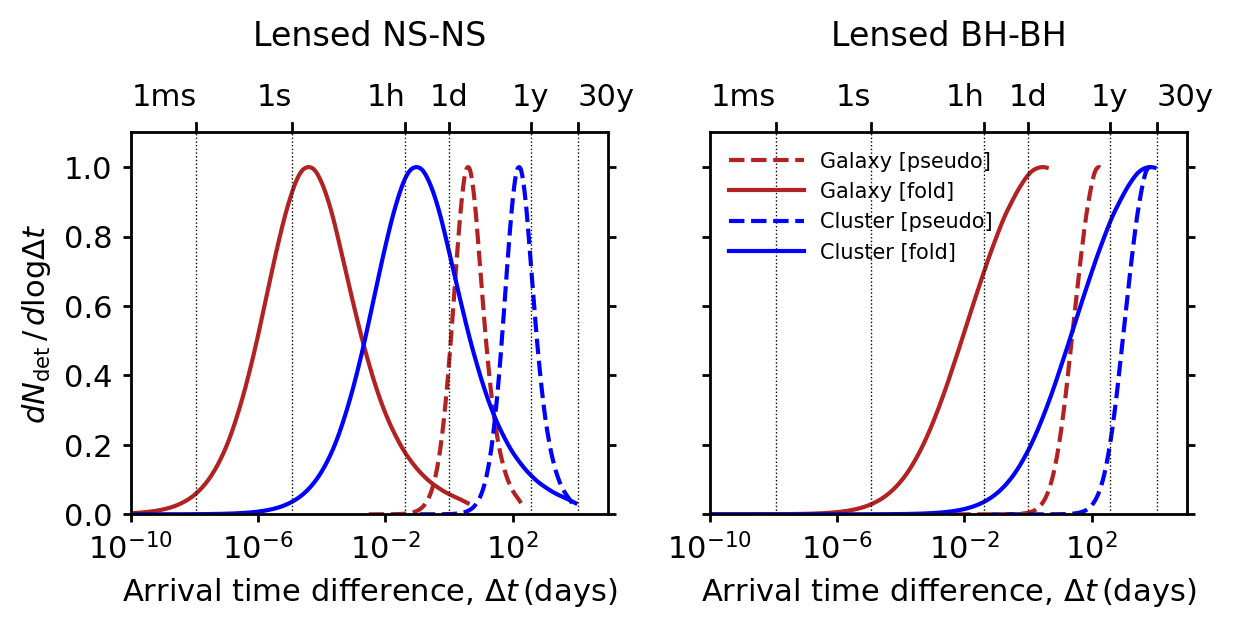}
  }
  \caption{{Lensed NS-NS mergers are better matched to the duration of GW detector runs than lensed BH-BH mergers because the arrival time difference between the GW signals from a lensed NS-NS image pair is predicted to be  sub-year (left panel), whilst the arrival time difference for lensed BH-BH image pairs is predicted to extend out to numerous years and potentially decades (right panel). } The panels show predicted arrival time difference based on our \emph{Baseline} model at the sensitivity forecast for the fifth GW run. Distributions are shown for typical galaxy-scale ($\thE=1\,\rm arcsec$, $\kE=0.5$, $\etaE=1$) and cluster-scale ($\thE=10\,\rm arcsec$, $\kE=0.8$, $\etaE=0.5$) lenses, as discussed in Section~\ref{sec:lensing:time}. }
  \label{fig:dtdist}
\end{figure*}

We use Equations~\ref{eqn:tpseudo} and \ref{eqn:tfold} to transform the predicted magnification distributions derived from our models (Figure~\ref{fig:mudist}) into predicted arrival time difference distributions ($\Dt_{\rm pseu}$ and $\Dt_{\rm fold}$) for typical galaxy-scale ($\thE=1\,\rm arcsec$, $\kE=0.5$, $\etaE=1$) and cluster-scale ($\thE=10\,\rm arcsec$, $\kE=0.8$, $\etaE=0.5$) lenses (Figure~\ref{fig:dtdist}). We adopt $\mathcal{D}=3.3\,\rm Gpc$, based on $\zs=1.6$ which corresponds to the peak of the predicted true distance ($D$) distribution discussed above, and $\zl=0.5$ which corresponds to the peak of the optical depth to strong lensing \citep{Robertson2020}. At fixed lens mass and structure the predicted $\Dt_{\rm fold}$ distributions are broader than the predicted $\Dt_{\rm pseu}$ distributions due to the stronger dependence of the former on lens magnification, $\mu$. Also, the predicted $\Dt_{\rm fold}$ distributions peak at smaller arrival time differences than the predicted $\Dt_{\rm pseu}$ distributions, again due to their respective dependence on $\mu$. 

{Gravitationally lensed BH-BH mergers suffer longer arrival time differences than gravitationally lensed NS-NS mergers because the former are typically less strongly magnified ($\mu\lesssim10$) than the latter ($\mu\simeq100$; Figure~\ref{fig:mudist}). Importantly, the predicted arrival time differences for lensed NS-NS mergers are dominated by sub-year timescales, whilst the arrival time differences for lensed BH-BH mergers can extend to numerous years, and in some cases decades. Detection of multiple signals from gravitationally lensed NS-NS mergers are therefore better matched to the length of GW detector runs than detection of multiple signals from lensed BH-BH mergers.}

\subsection{Comparison with previous work}\label{sec:pred:compare}

Our predictions for the rates of lensed GWs are broadly consistent with comparable earlier studies \citep[e.g.][]{Oguri2018,Oguri2019,Li2018,Ng2018,Wierda2021,Xu2022}. This is largely due to these studies adopting a broadly consistent description of the source populations and the lensing optical depth. However, these studies assumed that all lenses are galaxy-scale lenses and thus predicted the rate of multiply-imaged BH-BH (and in some cases NS-NS) mergers under that assumption. In contrast, we predict the rate of gravitationally magnified GWs agnostic to the mass of the dark matter halo that hosts the lens, and then interpret that prediction in the context of the predicted magnification distribution and the phenomenology of lenses as a function of their mass and structure (Section~\ref{sec:lensing:versus}). {For reasons discussed in Section~\ref{sec:lensing:odsl}, only a subset of the predicted lensed BH-BH population predicted in earlier work will be multiply-imaged due to the inefficiency of group/cluster-scale lenses at forming multiple images at $\mu<10$.} This underlines the importance of characterising the relationship between magnification and image multiplicity as a function of lens mass and structure with upcoming surveys such as the Vera C. Rubin Observatory's (Rubin's) Legacy Survey of Space and Time (LSST), ESA's \emph{Euclid} mission, and spectroscopic follow-up with ESO's 4MOST instrument. 

\subsection{Implications for discovery strategies}\label{sec:pred:implications}

Looking across the population of GW detections, it is clear from Figure~\ref{fig:contours} that essentially all GW detections within $\simeq1$ decade in $\tD$ of the horizon and with compact object masses in the range $2<\tm<100\,\rm M_\odot$ are candidate lensed detections. It is therefore particularly challenging to identify an individual detection at $\tm>5\,\rm M_\odot$ as a candidate lensed event because this overlaps with the BH mass function and the majority of BH-BH detections that are not expected to be lensed. {Moreover, selecting candidate lensed BH-BH mergers in the $\tm$-$\tD$ plane can be sensitive to population model choices
\citep[e.g.][]{Broadhurst2022}. In the $\tm$-$\tD$ plane, selecting candidate lensed NS-NS is $\simeq10\times$ more efficient (and less model dependent) than selecting candidate lensed BH-BH, because the majority of the former resides in the mass gap between NS and BH products, $2.5<\tm<5\,\rm M_\odot$ (Figure~\ref{fig:contours}). Indeed, objects have already been detected with non-zero probability in this region of the parameter space \citepalias{GWTC3}, some of which have been followed up electromagnetically \citep{Bianconi2022}.}

\begin{figure}
  \centerline{
    \includegraphics[width=0.8\hsize,angle=0]{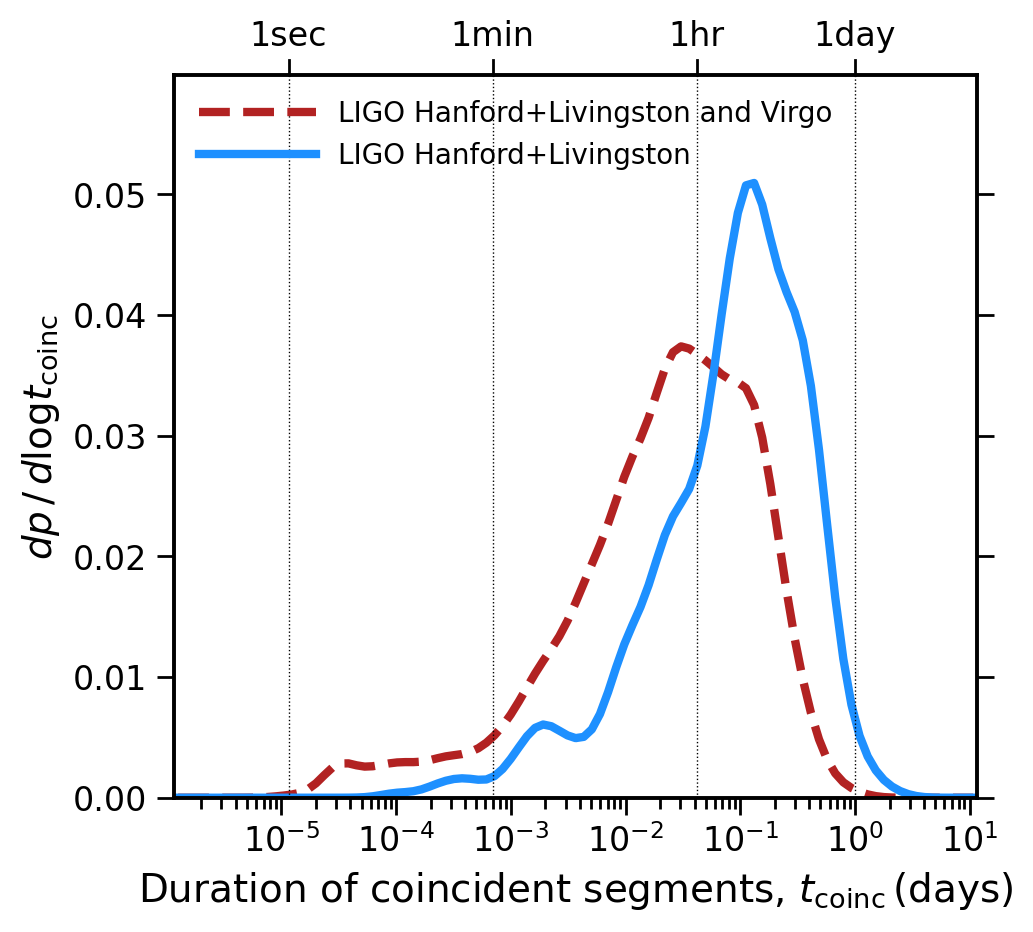}
  }
  \caption{Distribution of coincident segment durations from the third GW run, as discussed in Section~\ref{sec:pred:implications}, and based on data obtained from the Gravitational Wave Open Science Centre \citep{gwosc}.}
  \label{fig:tcoinc}
\end{figure}

Detection of two GW signals with consistent sky locations can also motivate the identification of candidate lensed objects \citep[e.g.][]{Dai2020,LVlens2021,Bianconi2022}. This strategy has two main disadvantages. Firstly, the significant sky localisation uncertainties of a growing catalogue of GW detections \citep{Petrov2021} are prone to generating false associations between detections. Secondly, each GW detector only collects data of a quality required for detection of binary compact object mergers for $\simeq70$ per cent of the duration of a run, and the typical coincident segment length is up to just a few hours. A segment is a period of time that a detector is collecting data and coincidence refers to overlap of segments at the different detectors (Figure~\ref{fig:tcoinc}). Moreover, the typical run length is a year with inter-run gaps of a similar duration. {Putting these practical constraints together, the distribution of coincident segment lengths points to overlapping sky localisations (and other parameter posteriors) of multiple detections within a few hours -- i.e.\ relatively short arrival time differences -- as being a good match between the operation of GW detectors and the predicted lensed GW population. This is an excellent match with lensed NS-NS mergers that produce image pairs formed by a fold caustic (Figure~\ref{fig:dtdist}). Clearly longer GW runs and within-run operational strategies that offset downtime of the different GW detectors, and thus minimize gaps between coincident segments will be advantageous for the detection of more than one image of a lensed GW.}

In summary, multiple arguments point to lensed NS-NS as having significant advantages for securing a robust and unambiguous lensed GW detection. These include short time delays, potential localisation to a lensed host galaxy, and selection of candidates in the $\tm$-$\tD$ plane in low latency that is less model dependent and $\simeq10\times$ more efficient than for candidate lensed BH-BH mergers. 

\section{Lensed kilonova counterparts: lightcurves and observing strategy}
\label{sec:observing}

We now discuss the prospects for localizing candidate lensed NS-NS mergers via the EM signal that accompanies them, focusing on prompt optical detection of kilonova counterparts because the detection of kilonovae is less dependent on viewing angle than the detection of SGRBs \citep{Metzger2018}. In Section~\ref{sec:knmodels} we give an overview of the kilonova models that we use, in Section~\ref{sec:lightcurves} we present and discuss predicted lightcurves of lensed kilonova counterparts, {in Section~\ref{sec:assumptions} we show that our reference lightcurve predictions are relatively insensitive to our assumptions,} and in Section~\ref{sec:obsstrategy} we discuss observing strategies required to detect the lensed kilonovae.

\subsection{Overview of kilonova models}\label{sec:knmodels}

\begin{table}
  \caption{Physical parameters of 170817-like and conservative kilonova models used for lightcurve predictions}
  \label{tab:knparams}
  \centering
  \begin{tabular}{lcc}
    \hline 
    Parameter & Conservative & 170817-like \cr
    \hline
    Chirp mass, $\mathcal{M}$ & $1.2\,\rm M_\odot$ & $1.188\,\rm M_\odot$\cr
    Mass ratio, $q$ & $0.9$ & $0.92$\cr
    Viewing angle & $60^\circ$ & $32^\circ$\cr
    Blue ejecta enhancement factor & 1 & 1.6 \cr    
    Cocoon opening angle & $0^\circ$ & $24^\circ$ \cr    
    \hline
  \end{tabular}
\end{table}

We use the recent kilonova light curve models from \citet{Nicholl2021}. These models, implemented within the Modular Open Source Fitter for Transients \citep[{\sc mosfit};][]{Guillochon2018}, include low, high and intermediate opacity ejecta components, representing the polar dynamical, tidal dynamical, and post-merger (wind) ejecta, respectively. The mass and velocity of each component is determined for a given binary configuration (chirp mass, mass ratio, and tidal deformability) using fits by \citet{Dietrich2017} and \citet{Coughlin2019} to numerical merger simulations. 

The luminosity is computed using the r-process decay heating rate \citep{Korobkin2012} with time-dependent thermalisation efficiency \citep{Barnes2016} and the \citet{Arnett1982} photon diffusion approximation. Further details of the r-process decay luminosity in \textsc{mosfit} are provided by \citet{Villar2017}. The \citeauthor{Nicholl2021} models also include a viewing angle dependence based on \citet{Darbha2020}, and an additional luminosity source from cooling emission of a `cocoon', shock-heated by a GRB jet escaping the ejecta, following \citet{Piro2018}. At a given time and wavelength, the apparent magnitude is calculated by summing blackbody spectral energy distributions for each component, adjusting for cosmological expansion at the specified redshift, and convolving with the transmission curve of the respective optical/near-infrared filters.

Given the diverse light curves that this model can produce, we elect here to study two representative cases that encapsulate the plausible range in luminosity. The optimistic or ``170817-like'' model simply takes the best fit parameters for GW170817, as determined by \citeauthor{Nicholl2021} (using the chirp mass inferred from its GW signal; \citealt{Abbott2017}), and evaluate this model at each redshift of interest. In the ``conservative'' case, we choose a chirp mass ($\mathcal{M}= 1.2$\,M$_\odot$) and mass ratio ($q = 0.9$) that are typical of known neutron star binaries that will merge within a Hubble time \citep{Farrow2019}. These values are similar to the 170817-like model, but three factors conspire to make this conservative case fainter (especially at blue wavelengths): we do not include blue ejecta enhancement by magnetic winds \citep{Metzger2018} or shock-heated GRB cocoon emission \citep{Piro2018}, and we rotate to a more representive viewing angle of 60$^\circ$ from the binary axis. The latter step further suppresses the blue emission (compared to the 170817-like model at $\approx 30^\circ$ off-axis), as less of the low opacity (lanthanide-poor) polar ejecta is within the line of sight for an edge-on observer. Both models are summarised in Table~\ref{tab:knparams}.

\subsection{Reference lightcurves of lensed kilonovae}
\label{sec:lightcurves}

\begin{figure*}
  \centerline{
    \includegraphics[width=0.7\hsize,angle=0]{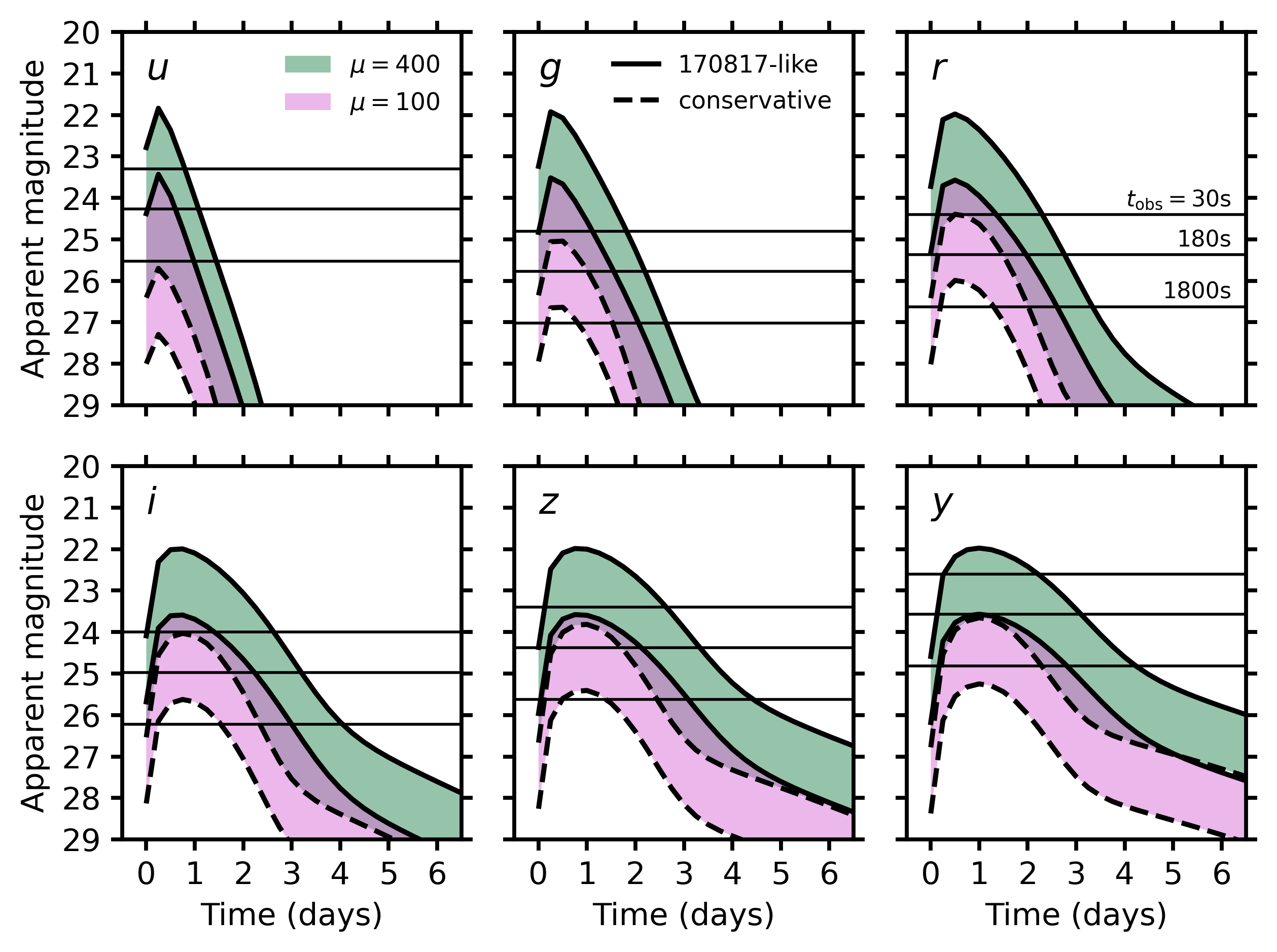}
  }
  \caption{Predicted lightcurves in the observer frame for lensed kilonova counterparts to lensed NS-NS mergers, based on the Vera C. Rubin Observatory's $ugrizy$-band transmission curves, with each panel labelled according to the respective filter. The solid and dashed curves show the predicted lightcurve for a kilonova counterpart that is similar to AT2017gfo counterpart to GW170817, and a redder more conservative model, respectively. The upper and lower solid curves show the predicted 170817-like lightcurve for lens magnifications representative of lensed NS-NS detections in the fourth and fifth GW runs respectively (Table~\ref{tab:peaks}). Similarly, the upper and lower dashed curves are representative of the fourth and fifth runs respectively. The green (upper) and pink (lower) shaded regions therefore indicate the predicted range of lightcurves for the fourth and fifth runs respectively. Thin horizontal lines show the depth expected to be reached by the Vera C. Rubin Observatory with LSSTCam with integration times of 30, 180, and 1800 seconds, as indicated in the top right panel. The time axis is relative to the time of the NS-NS merger in the observer frame. {Note that these lightcurves correspond to a lensed kilonova counterpart at a redshift of $z=1.6$, therefore the most sensitive filter ($r$-band) corresponds to a rest-frame wavelength of $\lambda_{\rm rest}\simeq230\,\rm nm$.}}
  \label{fig:lightcurves}
\end{figure*}

The observable lightcurve of a gravitationally magnified kilonova at a redshift of $z$ is related to the intrinsic lightcurve of the kilonova via the lens magnification and can be written in terms of apparent magnitudes in a given filter as:
\begin{eqnarray}
  \mfobs(z)&=&\mfint(z)-2.5\log\mu\cr
  &=&\mfint(z)-5\log(D/\tD).
  \label{eqn:mfobs}
\end{eqnarray}
A forward prediction of $\mfobs$ for a given kilonova model therefore requires a library of intrinsic lightcurves that describe $\mfint$ (in our case, based on \citeauthor{Nicholl2021}), and a choice of $\tD$ and $D$. To establish reference lightcurves for lensed kilonova counterparts to lensed NS-NS mergers, we base our forward predictions on our \emph{Baseline} NS-NS model, the peak of the distributions shown in Figure~\ref{fig:contours} for the fourth and fifth GW runs (Table~\ref{tab:peaks}), and {the transmission curves of the Rubin/LSST $ugrizy$-band filters}. 

\begin{table}
  \caption{Typical distances, redshifts and magnifications of lensed NS-NS and lensed kilonova counterparts as a function of GW detector sensitivity}
  \label{tab:peaks}
  \centering
  \begin{tabular}{lccccc}
    \hline 
    Parameter & O1 & O2 & O3 & O4 & O5 \cr
    \hline
    $\tD\,(\rm Mpc)$ & $270$ & $350$ & $500$ & $600$ & $1250$ \cr
    $D\,(\rm Gpc)$ & $12.1$ & $12.1$ & $12.1$ & $12.1$ & $12.1$ \cr
    $\mu$ & $2000$ & $1200$ & $600$ & $400$ & $100$ \cr
    $2.5\log\mu$ & $8.3$ & $7.7$ & $6.9$ & $6.5$ & $5.0$ \cr
    $\tz$ & $0.059$ & $0.075$ & $0.10$ & $0.13$ & $0.24$ \cr
    $z$ & $1.6$ & $1.6$ & $1.6$ & $1.6$ & $1.6$ \cr
    \hline
  \end{tabular}
\end{table}

{Our reference lightcurves reveal that lensed kilonova counterparts are faint and fade very quickly (Figure~\ref{fig:lightcurves}). In the fourth GW run ``170817-like'' and ``conservative'' lensed kilonovae will be detectable at $\rm AB\simeq22$ and $\rm AB\simeq24-25$ close to peak respectively, corresponding to a typical lens magnification for detectable lensed NS-NS mergers of $\mu\simeq400$. Because the sensitivity to faint or less magnified events will improve in the fifth GW run ($\tau^{S}\propto\mu^{-2}$; Equations~\ref{eqn:taumuth}~\&~\ref{eqn:taumus}), the typical lens  magnification will be lower, at $\mu\simeq100$. A typical lightcurve will thus peak $1.5$ magnitudes fainter than in the fourth run, at $\rm AB\simeq23.5$ and $\rm AB\simeq25.5-26.5$ for 170817-like and conservative kilonovae respectively. All the bright events will still be detectable -- the overall endeavour is not becoming more challenging -- but there will be many more frequent events at lower magnifications that can be explored.}

{Looking at $gri$-bands as the most sensitive of the optical/near-infrared filters from the ground, a clear signature of lensed kilonova counterparts is that they fade by $\gtrsim1$ magnitude per day within the first few days after peak (Figure~\ref{fig:lightcurves}). As such, they are the fastest fading class of transient sources, with a duration of $d\lesssim1$\,day (time taken to fade by a factor of two in flux from peak brightness in the observer's frame). They fade even faster than kilonovae that are not lensed, fast blue optical transients (FBOTs), and rapidly evolving transients (RETs), that all have $d\gtrsim2$\,days in the observer's frame \citep[e.g.][]{Perley2020}. This rapid evolution, and in particular its pace relative to kilonovae that are not lensed, is due to the filters probing the cosmologically time-dilated rest-frame UV emission from the distant lensed kilonovae ($\lambda_{\rm rest}\simeq150-300\,\rm nm$), as opposed to the rest-frame optical emission for nearby kilonovae that are not lensed ($\lambda_{\rm rest}\simeq400-700\,\rm nm$). Importantly, the pace of evolution of lensed kilonova counterparts is independent of the kilonova model, as can be seen by comparing solid and dashed curves in Figure~\ref{fig:lightcurves}. }

{In summary, lensed kilonova counterparts are distinctive because they fade faster than other optical transients studied to date. Also, in the mid/late-2020s they will typically be even fainter than the optical transients that Rubin/LSST is expecting to discover. Fast (spanning a few hours to a few days) and deep ($\rm AB\simeq25-26$) target of opportunity observations will therefore be required to detect them.}

\subsection{{Sensitivity of lightcurves to parameter choices}}
\label{sec:assumptions}

\begin{figure}
  \centerline{
    \includegraphics[width=\hsize,angle=0]{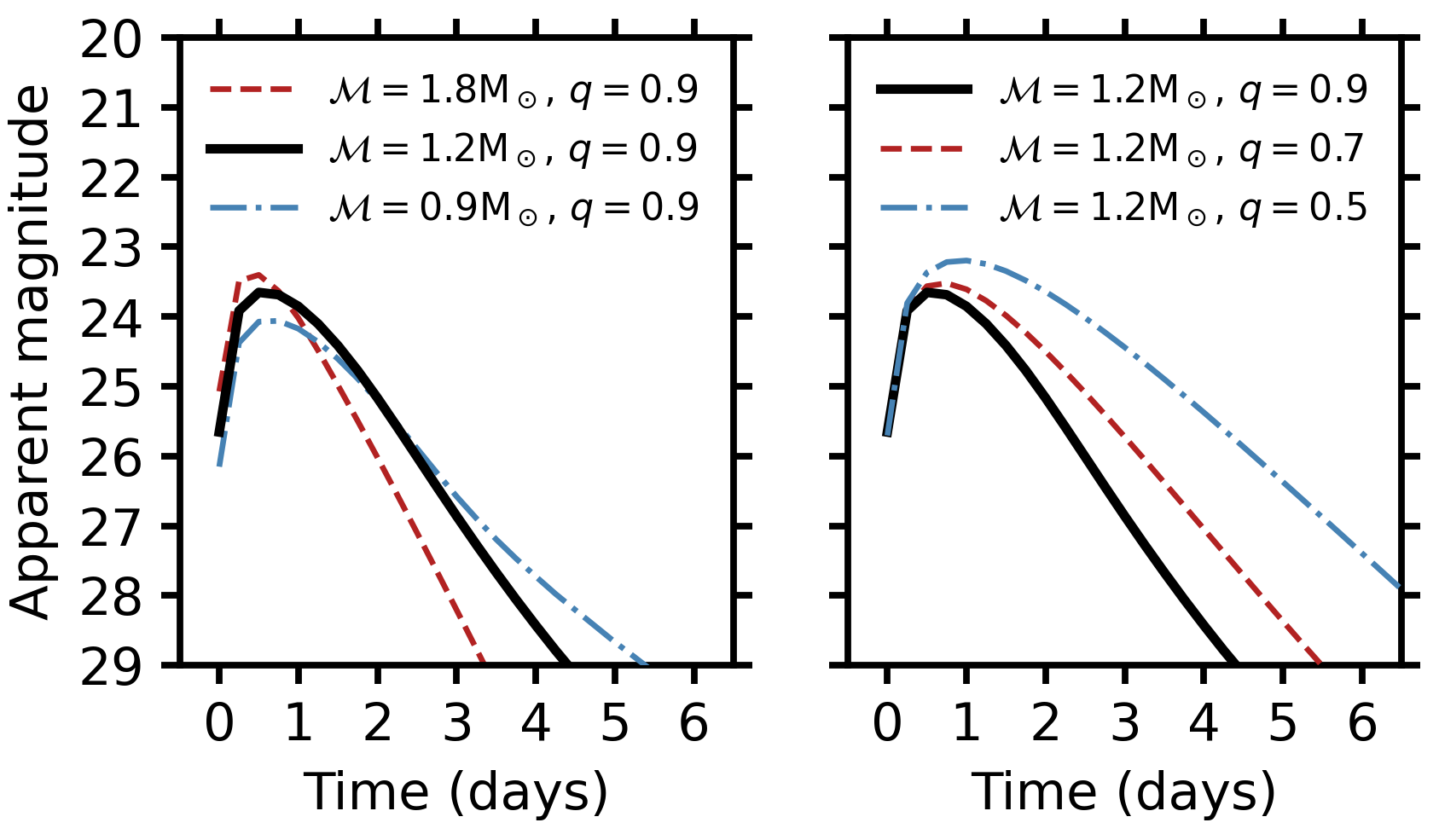}
  }
  \caption{Predicted {$r$-band} lightcurves for kilonova counterparts to lensed NS-NS mergers, concentrating on the fifth GW run, and demonstrating the sensitivity to the chirp mass (left) and mass ratio (right) of lensed NS-NS mergers with (as examples) cocoon opening angle of $30^\circ$ and a viewing angle of $60^\circ$. Kilonova counterparts to heavier lensed NS-NS mergers are brighter and fade more quickly than the counterparts to lighter lensed NS-NS mergers. Kilonova counterparts to less equal mass lensed NS-NS mergers are brighter and fade more slowly than the counterparts to more equal mass mergers. {However, the duration of the most extreme mass ratio systems ($q=0.5$) is $d<2\,\rm days$, and thus shorter than the general population of fast transients discussed in Section~\ref{sec:lightcurves}}. These characteristics are insensitive to the choice of opening and viewing angles. All other details of the figure are as stated in the caption to Figure~\ref{fig:lightcurves}. The black curves in both panels are identical.}
    \label{fig:qmchirp}
\end{figure}

The features of our reference lightcurves are robust to changing the chirp mass and mass ratio of the lensed NS-NS merger from which the lensed kilonova emission emanates. In Figure~\ref{fig:qmchirp} we show that varying chirp mass in the range $0.9<\mathcal{M}<1.8\,\rm M_\odot$ affects the apparent magnitude at peak brightness by $\ls0.5$ magnitudes, and that the  slope of the lightcurves within the first two days post NS-NS merger is not significantly altered by variations in chirp mass. We also show that more extreme mass ratio margers produce brighter kilonovae with lightcurves that fade more slowly than those with $q\simeq0.9-1$, under the assumption that the fraction of ejecta shock-heated by the GRB jet is independent of $q$. {However, in mergers with smaller mass ratios the bulk of the ejecta may be in the tidal plane, so this assumption may not hold if the GRB jet does not interact with such ejecta. Nevertheless, the most extreme mass ratio considered here ($q=0.5$) still has an observer frame duration of $d<2\,\rm days$, and thus our conclusion that lensed kilonovae are faster than all other transients is robust to variations in $q$.} Basing our reference lightcurves on $q=0.9$ is therefore conservative in the context of considering optical follow-up observing strategies in Section~\ref{sec:obsstrategy}.

Basing our reference lightcurves on the most probable distances and redshifts derived from our lens models is convenient, however the distributions shown in Figure~\ref{fig:contours} are broad. {We therefore examine the sensitivity of our reference lightcurves to these distributions}. Sensitivity to $\tD$ for a given $D$ is relatively straightforward to deal with, because the faintness of lensed kilonovae implies that it is only sensible to follow-up the candidate lensed NS-NS mergers with relatively small sky localization uncertainties. Better localized GW detections are located at smaller distances than worse localised detections \citep{Petrov2021}, which implies smaller $\tD$ for a given $D$, and thus larger lens magnification and brighter kilonovae. Therefore, when observing strategy considerations are folded in, basing the reference lightcurves on the peak distances and redshifts is conservative.

{Nevertheless}, $D$ is clearly unknown when identifying candidate lensed NS-NS mergers in low latency, when the LVK collaboration announces estimates of $\tD$, sky localisation posteriors, and the probability that one of the compact objects occupies the lower mass gap at $3<\tm<5\,\rm M_\odot$, $\pgap$. {We therefore} introduce a $k$-correction, $k(\tz,z,\lambda)$, as a convenient way to capture the uncertainty arising from unknown $D$. We write $\mfint(z)$ in terms of the apparent magnitudes the kilonova would have if it was located at $\tz$, $\mfint(\tz)$, a term to account for the inverse square law that recognises the impact of the difference between $\tD$ and $D$, and the $k$-correction that accounts for the different amounts of redshifting of the rest frame spectrum:
\begin{eqnarray}
  \mfint(z)&=&\mfint(\tz)+5\log(D/\tD)+k(\tz,z,\lambda),
  \label{eqn:mfint}
\end{eqnarray}
and then subsitute Equation~\ref{eqn:mfint} in to
Equation~\ref{eqn:mfobs} to obtain the following simple equation:
\begin{eqnarray}
  \mfobs(z)&=&\mfint(\tz)+k(\tz,z,\lambda).
  \label{eqn:kcorr}
\end{eqnarray}
Equation~\ref{eqn:kcorr} can be paraphrased as the lightcurve of the lensed kilonova counterpart to a lensed NS-NS merger is set by the lightcurve the object would have if it was really located at $\tD$ and the $k$-correction between $\tD$ and $D$. In other words, the inverse square law cancels with the lens magnification, and the unknown $D$ is encoded in the $k$-correction.

The $k$-corrections at redshifts of interest are not well constrained by existing data. At $z\simeq1-2$, most of the optical filters probe rest-frame UV emission, and even AT2017gfo was too faint for UV spectroscopy by the time observations were attempted ($t=5.5\,\rm days$; \citealt{Nicholl2017}). An estimate of the $k$-correction therefore requires models and assumptions about the kilonova physics. The \citet{Nicholl2021} models assume a multi-component blackbody spectral energy distribution, which is inevitably overly simplistic given the complicated atomic structures of the r-process elements that comprise the kilonova ejecta. However, even sophisticated radiative transfer models are likely to be unreliable at line-rich UV wavelengths, due to the difficulty of computing the atomic structure of all relevant species and the approximations used to get around this -- for example, using statistical properties of atoms rather than individual lines \citep{Kasen2017}, or specifying opacities rather than composition \citep{Bulla2019,Tanaka2020}.

Importantly, the \citet{Nicholl2021} model fits well the observed luminosity of AT2017gfo (effectively at $z=0$) in the \textit{Swift} UV bands (2000-3500\,\AA) at the most relevant time of $t\simeq1\,\rm day$. The observer-frame light curves of the 170817-like model should therefore be relatively trustworthy in all bands redder than $u$ at $z\simeq1-2$, but by $z\sim3$ the $k$-correction is not well constrained bluewards of the $i$-band. Since $z\simeq3$ is the redshift out to which lensed NS merger candidates can be identified through their detection in the lower mass gap, we can at least be confident that our predictions for a 170817-like kilonova are robust in the $i,z,y$-bands over the entire redshift range of interest (and in $g$ and $r$ over much of that range). For the conservative model, the $k$-correction is less secure at all wavelengths/redshifts, since this model has not been calibrated directly against an observed kilonova. As a guide, we therefore use the \citet{Nicholl2021} models to estimate $k$-corrections 1 day post merger in the $i$-band across the true redshift range $\tz<z<3$ for a source placed at $\tz=0.1$ in low latency, obtaining $-1\lesssim k\lesssim0$. This suggests that the $k$-corrections work in our favour and cause lensed kilonova to be a little brighter than if $k$-corrections are ignored. Therefore, despite the large uncertainties, the unknown true distance to lensed NS-NS appears unlikely to be a severe impediment to detecting their lensed kilonova counterparts.

\subsection{Outline observing strategy}\label{sec:obsstrategy}

{It is unlikely that a wide-field survey telescope will be observing the relevant sky region to the required depth when a candidate lensed NS-NS merger is detected by the GW detectors and announced as a public alert by the LVK collaboration. Target of Opportunity (ToO) optical observations will therefore need to be triggered in response to such a detection, with the upcoming Vera C.\ Rubin Observatory being the most powerful facility in the mid-2020s when detection rates approach one per year (Section~\ref{sec:pred:gw:rates}).}

{A sensitivity of $\simeq26-27$th magnitude is required to probe a broad range of kilonova models (Sections~\ref{sec:lightcurves}~\&~\ref{sec:assumptions}). As previously sketched by \citet{Smith2019lsst}, ToO observations that aim to detect lensed kilonova counterparts to candidate lensed NS-NS mergers (magnified in to the mass gap between NS and BH) need to be much deeper than those envisaged to follow-up NS-NS mergers that are not lensed \citep{Andreoni2021}. The lightcurves in Figure~\ref{fig:lightcurves} indicate that up to $\simeq30$\,minutes per pointing per epoch per filter will be required, depending on the expected magnification of the source. Such deep observations therefore motivate selecting well localized candidate lensed NS-NS mergers. Moreover, the combination of faint flux levels and distinctive short duration of lensed kilonovae motivates prioritising sensitivity in a single filter over attempting to measure colours from observations through multiple filters.} 

{Adopting nominal parameters of five Rubin/LSSTCam pointings ($9.6\,\rm degree^2$ per pointing), three epochs, one filter (preferably $r$-band), and $\lesssim30$\,minutes integration per pointing per epoch gives a ballpark estimate of $\lesssim7.5$\,hours of observing time per candidate lensed NS-NS -- i.e.\ $\lesssim0.25$ per cent of a year's observing time (assuming $3,000$ hours per year). This observing strategy will be sensitive to a broad range of blue and red kilonova models across the first two observer-frame days post GW detection. Two epochs on these timescales will aim to detect the rapid fading of the rest frame UV ($\lambda_{\rm rest}\simeq230\,\rm nm$ for the observer's $r$-band) lensed kilonova emission. Crucially, the cosmological time dilation and redshifting renders lensed kilonovae the fastest fading known transients at $\gtrsim1$\,magnitude per day. The third epoch (say) $\gtrsim4\,\rm days$ post GW detection would provide a post-hoc template image.}

Even a modest factor 2 reduction in the amount of observing time per candidate will help to increase the number of candidates that can be followed up and/or reduce the observing time requirement. In particular, release of additional mass information in low latency by the LVK collaboration would help to constrain our lensed kilonova lightcurve predictions. For example, the lightcurves of objects with lower mass ratios are brighter and fade slower than for higher mass ratios (Figure~\ref{fig:qmchirp}; note the mass ratio $q$ is lensing invariant and thus low latency posteriors on this parameter are straightforward to interpret). If available, low latency posteriors on $\tM$ would also help considerably, as highlighted by \cite{Bianconi2022} -- two of the three mass gap detections ($\pgap>0.94$) followed up in the third GW run were later identified as having chirp masses of $\tM\simeq8\,\rm M_\odot$. Low latency chirp mass posteriors would therefore enable a cleaner selection of candidates for expensive follow-up observations. {Such information will be similarly helpful for ToO follow-up observations with 4-m class telescopes including, CFHT/MegaCam and Blanco/DECam and dedicated GW follow-up facilities such as BlackGEM and GOTO. The latter facilities also typically have a very broad $VR$-band (or similar) filter, that effectively enhances ``light collecting power'' by a factor two over (say) the $r$-band. This can help to offset the smaller aperture size of these facilities when searching for faint fast lensed transients.}

In summary, rapid response deep ToO observations with the Vera C. Rubin Observatory are capable of detecting a lensed kilonova counterpart to a lensed NS-NS merger in the mid-2020s. Designing these observations to be useful for other science interests and gaining access to mass-related information in low latency will both go a long way to optimising the observations for mutual benefit across the community. GW detections in the mass gap are also scientifically compelling independent of the lensing interpretation. The observations outlined here therefore provide a physically well-motivated approach to exploring the mass gap electromagnetically, and will inform non-lensing interpretations of objects discovered in this region of parameter space. 

\section{Summary}\label{sec:summary}

{We have investigated gravitational lensing of GWs in the context of {the mass functions of compact object remnants of stellar evolution, and of strong gravitational lenses}. We developed a new analytic lens magnification-based approach to predicting the detectable rates of lensed GWs with the current generation of detectors, and transformed these predictions in to estimates of the arrival time difference between lensed images of the distant GW sources. This enabled us to compare different methods for selecting candidate lensed GWs, and to identify selecting candidate lensed NS-NS mergers that have been magnified in to the mass gap between NS and BHs as the most efficient method based on data that are publicly available immediately following the GW detection. Moreover, rapid selection of candidates lends itself to electromagnetic follow-up observations and localisation of candidate lensed NS-NS merger to a lensed host galaxy. This is the most direct way to test the interpretation that ``mass gap'' events have been gravitationally lensed, to shrink the sky localisation uncertainties of candidate lensed GWs to the angular scale of gravitational lensing, and thus to potentially achieve a secure first discovery of gravitational lensing of gravitational waves.} We therefore combined our lensing predictions with recent models of kilonova lightcurves from \citet{Nicholl2021} to predict the optical light curves of lensed kilonova counterparts to lensed NS-NS mergers in the filters that will be used by the upcoming Vera C.\ Rubin Observatory. Our main conclusions are as follows: 

\smallskip

\noindent{\bf Lensed GW detection becomes likely from 2023 onwards.}\newline The predicted rates of lensed BH-BH and lensed NS-NS mergers available to be detected will encroach on one per year during the fourth and fifth GW runs respectively. The predicted rates are relatively insensitive to the details of the mass functions we employed for the source populations, however significant uncertainty is due to the underlying local comoving merger rate density, with lensed rates of $\simeq1-10$ and $\simeq0.01-1$ per year for lensed BH-BH and lensed NS-NS mergers respectively from the mid-2020s.

\smallskip

\noindent{\bf Detection of lensed BH-BH mergers in low latency is challenging.}\newline Although more numerous than lensed NS-NS mergers, lensed BH-BH mergers are challenging to identify in low latency because their masses inferred assuming no magnification is at play ($\mu=1$) overlap with the population that is not lensed. In other words, in the mass-distance plane essentially every BH-BH detection is a \emph{candidate} lensed BH-BH. This is due to the steep slope of the BH mass function that has emerged from the catalogue of GW detections to date. This does not imply interpreting all BH-BH detections as being lensed -- the relative rate of lensed detections is $\simeq1{:}1000$ (Table~\ref{tab:relative}).

\smallskip

\noindent{\bf Multiple detections of lensed BH-BH mergers are challenging.}\newline Detection strategies based on identifying more than one lensed image of a BH-BH are also challenging. This is because lensed BH-BH mergers are predicted to be dominated by relatively low lens magnifications of $\mu\simeq2-10$. Consequently, the arrival time difference between lensed images of distant BH-BH mergers is typically long compared with the hour-long timescales on which GW detectors operate continuously without interruption. Moreover at these magnifications, lenses with relatively flat density profiles such as groups and clusters of galaxies are prone to only produce one lensed image. These issues reduce the efficiency of this selection method and motivate longer GW detector duty cycles {that maximize overlap between different GW detectors and detailed work on the relationship between magnification and multiplicity across the lens mass function}. 

\smallskip

\noindent{\bf Lensed NS-NS mergers are magnified in to the mass gap.}\newline Selection of candidate lensed NS-NS mergers in the mass-distance plane in low latency is relatively efficient because $>50$ per cent of them are predicted to {have been magnified in to the mass gap between NS and BH}. Selection of candidate lensed NS-NS mergers on the $\pgap$ measure that the LVK collaboration began releasing in low latency in their third run is therefore a useful tool. However, cleaner selection of candidates would be achieved if mass and/or mass ratio posteriors could be made available in low latency \citep{Bianconi2022}.

\smallskip

\noindent{\bf Lensed kilonova counterparts are detectable.}\newline {Combining our lensing models with \citet{Nicholl2021} kilonova models, we show that lensed kilonova counterparts peak at $\simeq22-26$th magnitude, depending on the level of lens magnification and whether a 170817-like or more conservative (redder) kilonova model is adopted. Whilst they are faint and fade quickly (the fastest fading transients, with a duration of $d<2\,\rm days$), the true redshift of the lensed kilonovae ($z\simeq1.6$) helps to stretch the light curves and make detection of the rest-frame UV emission feasible. We concentrate on deep target of opportunity observations with the upcoming Vera C.\ Rubin Observatory, and show that 3 epochs of $r$-band observations, spanning a sky localisation region of $\Omega_{90}\simeq50\,\rm degree^2$, would be sensitive to a broad range of kilonova physics, and thus enable detection of a lensed kilonova counterpart to a candidate lensed NS-NS. This equates to $\lesssim0.25$ per cent of a year's observing time per candidate lensed NS-NS. The most strongly magnified lensed kilonovae will also be detectable with smaller aperture telescopes, which will also benefit from using a broad $VR$-band (or similar) filter.}

\smallskip

\noindent{\bf Arrival time differences for lensed NS-NS/kilonovae are short.}\newline {Multiple detections of lensed NS-NS mergers are less challenging than for lensed BH-BH mergers because the former are more strongly magnified than the latter. In particular, the arrival time difference between lensed images of distant NS-NS mergers (sub-day) is typically comparable with the duty cycle GW detectors (see above). It is therefore plausible} that the second lensed image of a distant NS-NS merger is detected and the lensed kilonova counterpart starts to brighten during the optical follow-up observations that were triggered in response to the first image. This opens up the possibility of measuring the rest-frame UV lightcurve of a kilonova at a true redshift of $z\simeq1-2$ simultaneously with the NS-NS merger. 

\section*{Acknowledgments}

GPS dedicates this work to the memory of Derek Fry, and thanks Simon Birrer, Riccardo Buscicchio, Cressida Cleland, Tom Collett, Jeff Cooke, Ben Gompertz, Denis Martynov, Chris Moore, Sam Oates, Armin Rest, and Jielai Zhang for help and stimulating discussions. GPS acknowledges support from The Royal Society and the Leverhulme Trust. GPS and MB acknowledge support from the Science and Technology Facilities Council (grant numbers ST/N021702/1 and ST/S006141/1). GM acknowledges funding from the European Union’s Horizon 2020 research and innovation programme under the Marie Skłodowska-Curie grant agreement No MARACHAS - DLV-896778. MN is supported by the European Research Council (ERC) under the European Union’s Horizon 2020 research and innovation programme (grant agreement No.~948381) and by a Fellowship from the Alan Turing Institute. DR acknowledges a PhD studentship from the Science and Technology Facilities Council. MJ is supported by the United Kingdom Research and Innovation (UKRI) Future Leaders Fellowship (FLF), `Using Cosmic Beasts to uncover the Nature of Dark Matter' (grant number MR/S017216/1).

This research has made use of data from the Gravitational Wave Open Science Center (gw-openscience.org), a service of LIGO Laboratory, the LIGO Scientific Collaboration, the Virgo Collaboration, and KAGRA. LIGO Laboratory and Advanced LIGO are funded by the United States National Science Foundation (NSF) as well as the Science and Technology Facilities Council (STFC) of the United Kingdom, the Max-Planck-Society (MPS), and the State of Niedersachsen/Germany for support of the construction of Advanced LIGO and construction and operation of the GEO600 detector. Additional support for Advanced LIGO was provided by the Australian Research Council. Virgo is funded, through the European Gravitational Observatory (EGO), by the French Centre National de Recherche Scientifique (CNRS), the Italian Istituto Nazionale di Fisica Nucleare (INFN) and the Dutch Nikhef, with contributions by institutions from Belgium, Germany, Greece, Hungary, Ireland, Japan, Monaco, Poland, Portugal, Spain. The construction and operation of KAGRA are funded by Ministry of Education, Culture, Sports, Science and Technology (MEXT), and Japan Society for the Promotion of Science (JSPS), National Research Foundation (NRF) and Ministry of Science and ICT (MSIT) in Korea, Academia Sinica (AS) and the Ministry of Science and Technology (MoST) in Taiwan.

\section*{Data Availability}

The data presented in this article are available upon reasonable requests to the lead author.

\bibliographystyle{mnras}
\bibliography{main}

\appendix

\section{Gravitational time delay}\label{app:derivation}

We consider lenses that are approximately circular in projection on the sky, with a density profile local to a gravitationally lensed pair of images of the same source of:
\begin{equation}
    \kappa(x)=\langle\kappa\rangle\,x^{-\eta},
    \label{eqn:kappalocal}
\end{equation}
where $\kappa$ is defined in Equation~\ref{eqn:kappa_defn}, $x=\theta/\thE$ is the angle from the centre of the lens in units of the Einstein radius, $\langle\kappa\rangle$ is the density of the lens in the annulus that is bound by the location of the image pair, which for an axisymmetric lens is the convergence at the Einstein radius ($\kE$), and $0\leq\eta\leq1$. The arrival time difference between the images is conventionally phrased in terms of the image positions via $\Dtheta$, the width of the annulus referred to above, with the dependence on $\Dtheta$ differing between different types of image pairs. For singular lenses that form just two images (by virtue of their pseudo-caustic), $\Dt_{\rm pseu}\propto\Dtheta$ \cite[e.g.][]{Kochanek2002}, whilst for image pairs formed at fold and cusp catastrophes of non-singular lenses, $\Dt_{\rm fold}\propto\Dtheta^3$ and $\Dt_{\rm cusp}\propto\Dtheta^4$ respectively \cite[e.g.][]{Congdon2008}. Here we consider $\Dt_{\rm pseu}$ and $\Dt_{\rm fold}$, as they bracket the observed behaviour of lenses with measured time delays Section~\ref{sec:lensing:time}. In the following Sections we derive expressions for $\Dt_{\rm pseu}$ and $\Dt_{\rm fold}$ in terms of lens magnification, $\mu$, and lens structure, i.e.\ $\eta$ and $\langle\kappa\rangle$.

\subsection{Arrival time difference for the pseudo-caustic}
\label{app:pseudo}

\citet{Kochanek2002} showed that the arrival time difference for an image pair formed by a pseudo-caustic is given by
\begin{equation}
    \Dt_{\rm pseu}=2\Dt_{\rm SIS}\left[1-\langle\kappa\rangle-\frac{1+(1-\eta)\langle\kappa\rangle}{12}\left(\frac{\Dtheta}{\langle\theta\rangle}\right)^2+\mathcal{O}\left(\left(\frac{\Dtheta}{\langle\theta\rangle}\right)^4\right)\right],
    \label{eqn:dtkochanek}
\end{equation}
where $\Dt_{\rm SIS}$ is the arrival time difference between two images formed by a singular isothermal sphere (SIS lens). Since $1\geq\eta\geq0$ and $0.5\leq\langle\kappa\rangle\leq1$ (Figure~\ref{fig:foxetal+odsl}), it is clear that the $(\Dtheta/\langle\theta\rangle)^2$ term contributes no more than $\simeq10$ per cent to $\Dt_{\rm pseu}$, even for $\Dtheta/\langle\theta\rangle\simeq1$. Therefore, $\Dt_{\rm pseu}$ is mainly influenced by the density of the lens, $\langle\kappa\rangle$:
\begin{equation}
    \Dt_{\rm pseu}\simeq2\,\Dt_{\rm SIS}\,\bigl[1-\langle\kappa\rangle\bigr].
\end{equation}
Therefore, $\Dt_{\rm pseu}\rightarrow0$ for a flat density profile ($\eta\rightarrow0$, $\langle\kappa\rangle\rightarrow1$), and we recover $\Dt_{\rm pseu}=\Dt_{\rm SIS}$ for an isothermal lens ($\eta=1$, $\langle\kappa\rangle=0.5$). 

The arrival time difference for a SIS lens is typically written in terms of the positions of the two images, $\thetap$ and $\thetam$:
\begin{equation}
    c\,\Dt_{\rm SIS}=\frac{\mathcal{D}}{2}\bigl[\thetap^2-\thetam^2\bigr]=\mathcal{D}\,\langle\theta\rangle\,\Dtheta,
    \label{eqn:dtsis}
\end{equation}
where $\mathcal{D}$ is defined adjacent to Equation~\ref{eqn:ct}. We now use Equation~\ref{eqn:musis} to re-write Equation~\ref{eqn:dtsis} in terms of magnification. Specifically, the sum of the lens magnifications suffered by the image pair is $\mup=2\thE/\beta=2\langle\theta\rangle/\beta$ where $\beta$ is the source position, and $\theta_{\pm}=\beta\pm\thE$, therefore $\Dtheta=2\beta$. Putting this together, we obtain
\begin{equation}
    c\,\Dt_{\rm SIS}=\frac{4\,\mD\,\langle\theta\rangle^2}{\mup},
    \label{eqn:dtsismup}
\end{equation}
and thus
\begin{equation}
    \Dt_{\rm pseu}\simeq\frac{8\mathcal{D}\,\bigl[1-\langle\kappa\rangle\bigr]\,\langle\theta\rangle^2}{c\,\mup},
    \label{eqn:dtpseudo}
\end{equation}
We also write Equation~\ref{eqn:dtpseudo} as a convenient scaling relation:
\begin{equation}
    \frac{\Dt_{\rm pseu}}{92\,\rm d}=\left[\frac{\thE}{1''}\right]^2\,\bigg[\frac{\mup}{4}\bigg]^{-1}\,\bigg[\frac{1-\kE}{0.5}\bigg]\,\left[\frac{\mathcal{D}}{3.3\,\rm Gpc}\right],
\end{equation}
where $\mathcal{D}=3.3\,\rm Gpc$ corresponds to $\zl=0.5$ and $\zs=1.6$, and we have approximated $\kE=\langle\kappa\rangle$ and $\thE=\langle\theta\rangle$. 

\subsection{Image pairs formed by fold catastrophes}\label{app:fold}

We start from \citeauthor{Schneider1992}'s (\citeyear{Schneider1992}) expression for the arrival time difference (their Equation 6.21a) between two fold images:
\begin{equation}
  c\,\Dt_{\rm fold}=\frac{\mathcal{D}\,\Dtheta^2}{6\,\mup\left|\tau_{11}^{(0)}\right|},
  \label{eqn:dtschneider}
\end{equation}
where $\tau_{11}^{(0)}$ is the partial second order derivative of the Fermat potential evaluated at the {critical} point of the lens mapping, i.e.\ where $\det(\mathcal{A})=0$) at the mid-point between the images. We then use \citeauthor{Schneider1992}'s Equation 6.20 to eliminate $\Dtheta$, to obtain
\begin{equation}
    c\,\Dt_{\rm fold}=\frac{2\,\mathcal{D}\,}{3\,\mup^3\,\left|\tau_{11}^{(0)}\right|^3\,\left|\tau_{222}^{(0)}\right|^2}.
    \label{eqn:dtfold}
\end{equation}

Next, we address the dependence of $\Dt_{\rm fold}$ on the structure of the lens. Taking $\tau_{11}$ first, we obtain from Equations~\ref{eqn:fermat}~\&~\ref{eqn:jacobian}:
\begin{equation}
  \tau_{11}=1-\psi_{11}=\mathcal{A}_{11},
\end{equation}
where the subscripts on $\mathcal{A}$ denote the elements of the Jacobian matrix. Following \citeauthor{Schneider1992}, we can write this as \begin{equation}
  \tau_{11}=2(1-\kappa),
\end{equation}
and then evaluate $\tau_{11}$ at the mid-point of the image pair to obtain:
\begin{equation}
  \left|\tau_{11}^{(0)}\right|=2\,\big|1-\langle\kappa\rangle\big|.
  \label{eqn:tau11}
\end{equation}

Turning to $\tau_{222}$, we differentiate Equation~\ref{eqn:fermat} to obtain:
\begin{equation}
  \tau_{222}=-\psi_{222},
\end{equation}
which shows, in line with many previous works
\citep[e.g.][]{Goldberg2005,Bacon2006,Lasky2009}, that second order lensing (flexion) depends on derivatives of the convergence $\kappa$. We therefore adopt $\tau_{222}=\partial\kappa/\partial\theta$ and write $\left|\tau_{222}^{(0)}\right|$ as:
\begin{equation}
  \left|\tau_{222}^{(0)}\right|=\left|\frac{d\kappa}{d\theta}\right|_{\langle\theta\rangle},
  \label{eqn:tau222a}
\end{equation}
where the subscript denotes evaluation of the derivative at $\theta=\langle\theta\rangle$. Substituting Equation~\ref{eqn:kappalocal} in to Equation~\ref{eqn:tau222a} then yields
\begin{equation}
  \left|\tau_{222}^{(0)}\right|=\frac{\eta\,\langle\kappa\rangle}{\langle\theta\rangle}.
  \label{eqn:tau222b}
\end{equation}

Substituting Equations~\ref{eqn:tau11}~and~\ref{eqn:tau222b} into Equation~\ref{eqn:dtfold} we obtain:
\begin{equation}
  \Dt_{\rm fold}=\myfrac[2pt][2pt]{\mathcal{D}\,\langle\theta\rangle^2}{12\,c\,\mup^3\,\eta^2\,\langle\kappa\rangle^2\,\big[1-\langle\kappa\rangle\big]^3},
\end{equation}
which can be phrased as a convenient scaling relation:
\begin{equation}
    \frac{\Dt_{\rm fold}}{3.9\,\rm days}=\left[\myfrac[0pt][1pt]{\thE}{1''}\right]^2\left[\myfrac[0pt][1pt]{\mup}{4}\right]^{-3}\left[\myfrac[0pt][1pt]{\etaE}{1}\right]^{-2}\left[\myfrac[0pt][1pt]{\kE}{0.5}\right]^{-2}\left[\myfrac[0pt][1pt]{1-\kE}{0.5}\right]^{-3}\left[\myfrac[0pt][1pt]{\mathcal{D}}{\rm 3.3\,Gpc}\right].
\end{equation}

\bsp
\label{lastpage}
\end{document}